\documentclass[reprint,twocolumn,superscriptaddress,secnumarabic,amssymb, nobibnotes, aps, prl]{revtex4-1}

\usepackage{amsmath}    
\usepackage{graphicx}    
\usepackage{verbatim}    
\usepackage{color}           
\usepackage{subfigure}   
\usepackage{hyperref}    
\usepackage{braket}
\usepackage{physics}
\usepackage{xfrac}
\usepackage{xspace}
\newcommand{\amt}{$\alpha$-MnTe\xspace}
\newcommand{\bk}{\mathbf{k}}

\newcommand{\bbm}{\mathbf{m}}
\newcommand{\bb}{\mathbf{b}}
\newcommand{\bL}{\mathbf{L}}
\newcommand{\cb}{[11\bar{2}0]}
\newcommand{\ca}{[1\bar{1}00]}
\def\bL{\mathbf{L}}
\def\hp{\hat{p}}
\def\hd{\hat{d}}
\def\hH{\hat{H}}
\def\hn{\hat{n}}
\def\hT{\hat{T}}

\begin{document}
\title{X-ray Magnetic Circular Dichroism in Altermagnetic \amt}

\author{A.~Hariki}
\affiliation{Department of Physics and Electronics, Graduate School of Engineering,
Osaka Metropolitan University, 1-1 Gakuen-cho, Nakaku, Sakai, Osaka 599-8531, Japan}
\author{A.~Dal~Din}
\affiliation{School of Physics and Astronomy, University of Nottingham, Nottingham NG7 2RD, United Kingdom}
\author{O.~J.~Amin}
\affiliation{School of Physics and Astronomy, University of Nottingham, Nottingham NG7 2RD, United Kingdom}
\author{T.~Yamaguchi}
\affiliation{Department of Physics and Electronics, Graduate School of Engineering,
Osaka Metropolitan University, 1-1 Gakuen-cho, Nakaku, Sakai, Osaka 599-8531, Japan}
\author{A.~Badura}
\affiliation{Institute of Physics, Czech Academy of Sciences, Cukrovarnick\'{a} 10, 162 00 Praha 6 Czech Republic} 
\author{D.~Kriegner}
\affiliation{Institute of Physics, Czech Academy of Sciences, Cukrovarnick\'{a} 10, 162 00 Praha 6 Czech Republic} 
\author{K.~W.~Edmonds}
\affiliation{School of Physics and Astronomy, University of Nottingham, Nottingham NG7 2RD, United Kingdom}
\author{R.~P.~Campion}
\affiliation{School of Physics and Astronomy, University of Nottingham, Nottingham NG7 2RD, United Kingdom}
\author{P.~Wadley}
\affiliation{School of Physics and Astronomy, University of Nottingham, Nottingham NG7 2RD, United Kingdom}
\author{D.~Backes}
\affiliation{Diamond Light Source, Chilton OX11 0DE, United Kingdom}
\author{L.~S.~I.~Veiga}
\affiliation{Diamond Light Source, Chilton OX11 0DE, United Kingdom}

\author{S.~S.~Dhesi}
\affiliation{Diamond Light Source, Chilton OX11 0DE, United Kingdom}
\author{G.~Springholz}
\affiliation{Institute of Semiconductor and Solid State Physics,
Johannes Kepler University Linz, Altenbergerstr. 69, 4040 Linz, Austria}
\author{L.~\v{S}mejkal}
\affiliation{Institut f\"ur Physik, Johannes Gutenberg Universit\"at Mainz, D-55099 Mainz, Germany}
\affiliation{Institute of Physics, Czech Academy of Sciences, Cukrovarnick\'{a} 10, 162 00 Praha 6 Czech Republic}
\author{K.~V\'yborn\'y}
\affiliation{Institute of Physics, Czech Academy of Sciences, Cukrovarnick\'{a} 10, 162 00 Praha 6 Czech Republic}
\author{T.~Jungwirth}
\affiliation{Institute of Physics, Czech Academy of Sciences, Cukrovarnick\'{a} 10, 162 00 Praha 6 Czech Republic}
\affiliation{School of Physics and Astronomy, University of Nottingham, Nottingham NG7 2RD, United Kingdom}
\author{J.~Kune\v{s}}
\affiliation{Institute for Solid State Physics, TU Wien, 1040 Vienna, Austria}
\affiliation{Department of Condensed Matter Physics, Faculty of
  Science, Masaryk University, Kotl\'a\v{r}sk\'a 2, 611 37 Brno,
  Czechia}

\begin{abstract}
Altermagnetism is a recently identified magnetic symmetry class combining characteristics of conventional collinear ferromagnets and antiferromagnets, that were regarded as mutually exclusive, and enabling phenomena and functionalities unparalleled in either of the two traditional elementary magnetic classes. In this work we use symmetry, {\em ab initio} theory and experiments to explore X-ray magnetic circular dichroism (XMCD) in the altermagnetic class. As a representative material for our XMCD study we choose $\alpha$-MnTe with compensated antiparallel magnetic order in which an anomalous Hall effect has been already demonstrated. We predict and experimentally confirm a characteristic XMCD lineshape for compensated moments lying in a plane perpendicular to the light propagation vector. Our results highlight the distinct phenomenology in altermagnets of this time-reversal symmetry breaking response, and its potential utility for element-specific spectroscopy and microscopy.

\end{abstract}

\maketitle

Recent theoretical studies have identified magnetic crystals with unconventional characteristics. On one hand, the crystal symmetries generate a compensated antiparallel magnetic order. On the other hand, they enable time-reversal ($\cal{T}$) symmetry breaking linear responses, such as the anomalous Hall effect (AHE) \cite{Smejkal22b,Smejkal22a,Smejkal20,Samanta20,Naka20,Reichlova21,Hayami21,Mazin21,Betancourt23,Naka22} or charge-spin conversion effects \cite{Smejkal22a,Naka19,Gonzalez-Hernandez21,Naka21,Ma21,Smejkal22c,Smejkal22}, and strongly spin-polarized electronic band structures \cite{Smejkal22a,Smejkal20,Ahn19,Hayami19,Yuan20,Yuan21,Hayami20,Reichlova21,Smejkal22,Mazin21,Liu22,Jian23}. Theoretical predictions have gone beyond the customary notion that these phenomena originate from finite magnetization in ferromagnets, or from non-collinear magnetic order on certain lattices \cite{Smejkal22b}. {\em Ab initio} calculations in collinear magnets have identified large non-relativistic spin splittings in the band structure (reaching $\sim$~eV in  RuO$_2$ or MnTe 
\cite{Smejkal20,Ahn19,Betancourt23}) 
which have subsequently been confirmed by photoemission spectroscopy~\cite{Krempasky23}.
Together with experimental observation of AHE in RuO$_2$ or MnTe \cite{Feng22,Betancourt23} these are hallmarks of altermagnetism~\cite{Smejkal22a}. Also,  predictions of  strong spin currents, opening the prospect of robust writing and readout mechanisms in stray-field-free ultra-fast memory devices \cite{Smejkal22a,Gonzalez-Hernandez21,Shao21,Smejkal22c,Smejkal22},  have been supported by initial  experiments in RuO$_2$ \cite{Bose22,Bai22,Karube22}.

\begin{figure}
\includegraphics[width=\columnwidth]{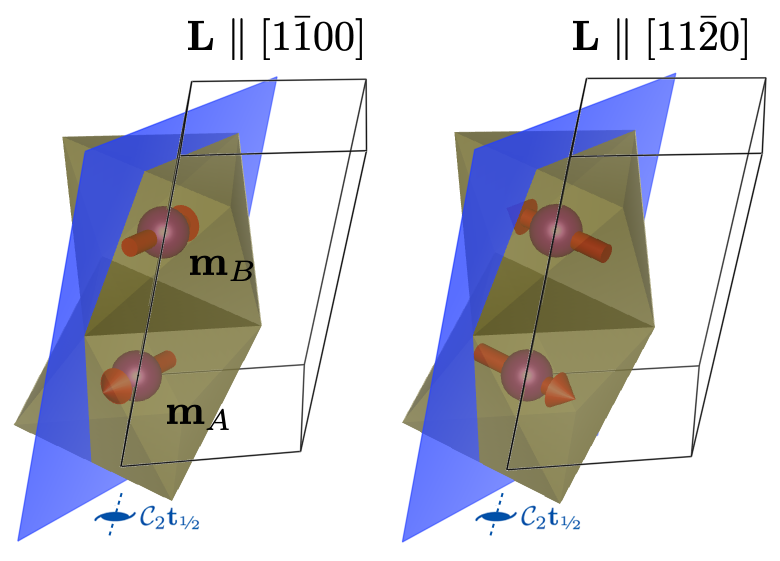}
\caption{Mn moments $\bbm_A$ and $\bbm_B$ in \amt structure (with Te octahedra)
for the studied N\'eel vector $\bL$ orientations. 
The mirror plane $\mathcal{M}$ discussed in the text is marked in
blue. While in the right panel $\mathcal{M}$ is an element of the magnetic symmetry group, 
in the left panel it is $\mathcal{MT}$ which leaves the system invariant.
}
\label{fig:geometry}
\end{figure}
In this letter we study the X-ray magnetic circular dichroism (XMCD) and demonstrate that, apart from the ferromagnetic or non-collinear order \cite{Wimmer19,Kimata2021,vdLaan2021}, it can also originate from the collinear order in altermagnets. Both AHE, the transverse electric-current response to an applied electric bias, and XMCD, the difference between absorption of right and left circularly polarized X-rays $F_+-F_-$, are given by the Hall vector ${\bf h} = (\sigma^a_{zy}, \sigma^a_{xz}, \sigma^a_{yx})$; here $\sigma^a_{ij}=-\sigma^a_{ji}$ are the antisymmetric components of the (frequency dependent) conductivity tensor \cite{Smejkal22b,Wimmer19}.  
On this level, AHE and XMCD are, therefore, governed by the same symmetry principles. 

Before 
discussing XMCD in \amt, we summarize key symmetry considerations. 
The non-relativistic 
symmetries 
consist of direct products of transformations in the decoupled real (orbital) and spin spaces and allow to distinguish three classes~\cite{Smejkal22a,Smejkal22} among collinear magnets: first,  ferromagnets (ferrimagnets) have one spin lattice or several spin sublattices not connected by any symmetry transformation. Next, an antiferromagnetic class has the opposite spin sublattices connected by a real-space translation or inversion~\cite{Smejkal2017,Elmers2020}. Finally,  an altermagnetic class has the opposite spin sublattices connected by a real-space rotation, 
but not connected by a translation or inversion. Unlike the 
ferromagnets with a net non-relativistic magnetization, 
and unlike the antiferromagnets with 
spin-unpolarized $\cal{T}$-invariant bands, altermagnets have zero non-relativistic net magnetization combined with spin-polarized bands that  break $\cal{T}$-symmetry. This suggests
the presence of  $\cal{T}$-symmetry breaking responses, analogous to ferromagnets,
but with qualitatively different phenomenology in the intrinsically anisotropic (even-parity $d$, $g$ or $i$-wave) altermagnets  \cite{Ahn19,Smejkal22a,Smejkal22}.

In the absence of the relativistic spin-orbit coupling (SOC), collinear magnets  have the combined ${\cal C}_2^S\cal{T}$ symmetry,
where $C_2^S$ is the spin-space rotation around an axis perpendicular to the ordered moments.
This symmetry together with the invariance of the Hall vector ${\bf h}$ under an arbitrary spin-space rotation make ${\bf h}$ vanish in the absence of SOC.
In relativistic physics, the real space and the spin space are coupled
eliminating pure spin rotations  
as well as the ${\cal C}_2^S\cal{T}$ spin-space symmetry.
In ferromagnets, the  Hall vector  is 
always allowed in the presence of SOC because both the magnetization and the Hall vector transform as $\cal{T}$-odd axial vectors \cite{Smejkal22}. 
\begin{figure}[t]
\includegraphics[width=\columnwidth]{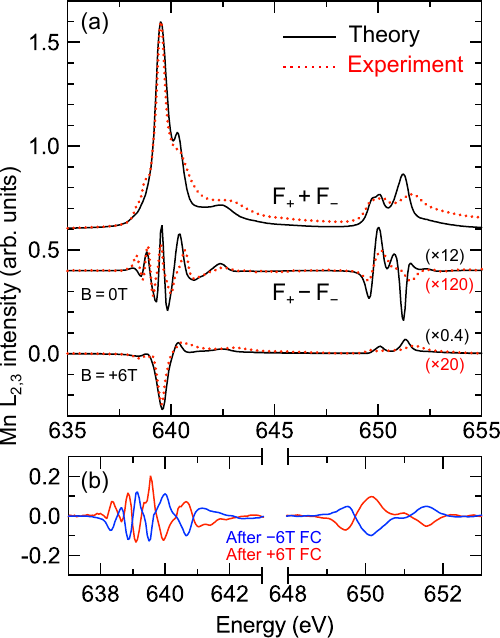}
    \caption{a) Mn $L_{2,3}$-edge spectra in theory (black) and experiment (red dotted) for
    light propagating along the $c$-axis. 
    Top: XAS. Middle: XMCD in zero field after cooling in 6~T.  
    Bottom: measured XMCD in the applied field of 6~T (red dotted), calculated XMCD for $\bk$ parallel
    to the magnetic moment scaled by $m_{6\text{T}}/m_\text{sat}\approx 1/50$ (black). b) XMCD after cooling in the opposite fields.}
\label{fig:xas}
\end{figure}
In altermagnets, the symmetry lowering by SOC also enables the presence of the  Hall vector, but not necessarily for all directions of the N\'eel vector \cite{Smejkal22}. 

\begin{figure}[t]
\includegraphics[width=0.57\columnwidth]{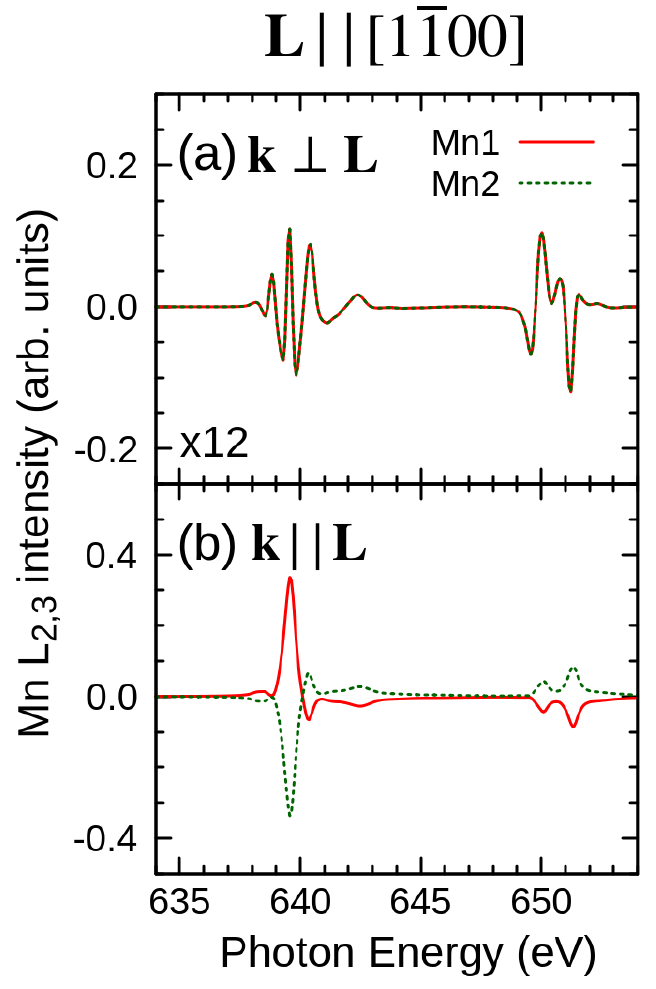}
\includegraphics[width=0.41\columnwidth]{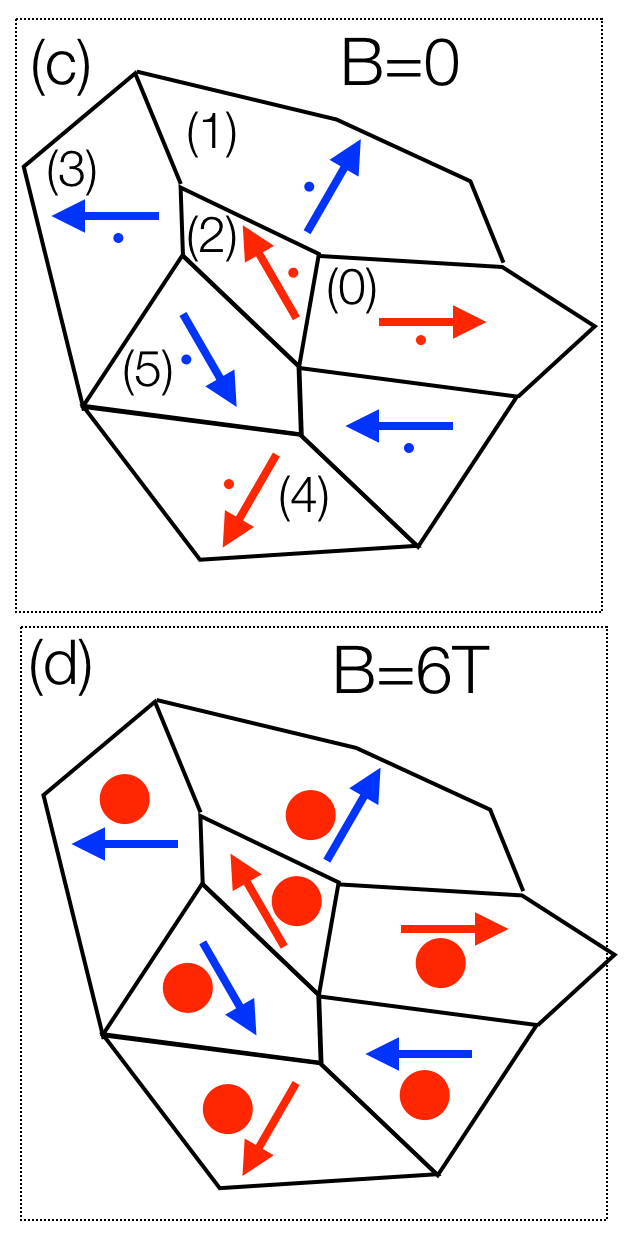}
    \caption{The Mn site-resolved contributions to XMCD calculated by the LDA+DMFT AIM for the N\'eel vector $\bL \parallel\ca$ and the light propagation vector $\bk \perp {\bf L}$ (a) and $\bk \parallel {\bf L}$ (b). (c) A cartoon view along the $c$-axis of the possible domain structure with the six easy-axis orientations of $\bL$. The domains with even labels (red) contribute $\Delta F(\omega)$ with positive prefactor, the odd ones (blue) with negative prefactor. The red and blue dots indicate the positive and negative orientation of the out-of-plane magnetization $\bbm$. (d) 
    The out-of-plane canting of the moments in 6~T applied field does not strongly depend on the domain's $\bL$
    (The domain sizes and shapes were chosen randomly and are not
    intended to have physical meaning.)}
\label{fig:xmcd}
\end{figure}

We now proceed to the analysis of the XMCD in MnTe. A schematic crystal of \amt with NiAs structure (crystallographic space group $P6_3/mmc$ \#194 \cite{villars}) is shown in Fig.~\ref{fig:geometry}. The magnetic moments on Mn have a parallel alignment within the $c$-planes and an antiparallel alignment between the planes. The face-sharing Te octahedra surrounding the Mn atoms break the translation or inversion symmetry connecting the opposite-spin sublattices, but allow a non-symmorphic six-fold screw rotation $\mathcal{C}_{6}^Rt_{\sfrac{1}{2}}$ to connect them \cite{Smejkal22,Betancourt23} rendering \amt an altermagnet. The magnetocrystalline anisotropy together with the crystal symmetry define six easy-axis orientations of the N\'eel vector $\bL=\bbm_A-\bbm_B$, with $\bbm_A$ and $\bbm_B$ being the sublattice moments, which lie in the $c$-plane~\cite{Kriegner2017,Betancourt23}. The easy axes form two triples $[1\bar{1}00]$, $[01\bar{1}0]$, $[\bar{1}010]$ and $[\bar{1}100]$, $[0\bar{1}10]$, $[10\bar{1}0]$. We label them
with even and odd integers, respectively, in
Fig.~\ref{fig:xmcd}c. The states within the triples are related by ${\cal C}_3$ rotations
about the $c$-axis, while the mapping between the triples is provided by the time reversal $\mathcal{T}$. Relativistic symmetry for 
these orientations of $\bL$ allows finite net magnetization $\bbm_A+\bbm_B=\bbm\parallel c$, due to canting of $\bbm_A$ and $\bbm_B$ towards the $c$-axis. The symmetry implies that $\bbm$ has the same sign within the triples, but opposite between them.
The magnitude $|{\bf m}|$ was experimentally determined to be extremely small (below $10^{-4}\,\mu_B$ per Mn atom in bulk \amt~\cite{Kluczyk23} and around $10^{-3}\,\mu_B$ in the present thin-film epilayers).



The X-ray absorption spectrum (XAS) and XMCD at the Mn $L_{2,3}$ edge are calculated
using a dynamical mean-field theory (DMFT) + Anderson impurity model (AIM) approach of Refs.~\onlinecite{Hariki2017,Hariki18,Hariki20}.
Starting with a density functional calculation using Wien2k~\cite{wien2k} we construct a multi-band Hubbard model~\cite{wien2wannier,wannier90} spanning the Mn $3d$ and Te $5p$ bands. The Coulomb interaction within the Mn $d$-shell is parametrized by $U=$~5.0~eV and $J=$~0.86~eV, and the double-counting correction $\mu_{\text{dc}}=$~22.5~eV. DMFT with these parameters reproduces well the valence-band both in experiment~\cite{Sato99} and theory~\cite{pfjr:23} as well as Mn $2p$ core-level X-ray photoemission spectra~\cite{Jian10}, see SM~\cite{sm}.

In Fig.~\ref{fig:xas}a we compare the calculated XAS and XMCD for the light propagating along the $c$-axis with the experiment. The measurements were performed on an epitaxial \amt\ 30~nm thick film grown on a InP (111) substrate by molecular beam epitaxy as described in Ref.~\onlinecite{Kriegner2017}. Samples were transferred in vacuum from the growth chamber to beamline I06 at Diamond Light Source in a vacuum suitcase. The XAS measurements with circularly polarized light propagating along the $c$-axis were performed at 140~K using the total electron yield detection.
The sample is in zero magnetic field after cooling from above the N\'eel temperature in a 6~T magnetic field applied along the $c$-axis. This procedure results in preferential population of domains with $\bL=\ca$, $[01\bar{1}0]$ and $[\bar{1}010]$, over domains with $-\bL$. Cooling in a reversed magnetic field results in the sign reversal of the XMCD signal shown in Fig.~\ref{fig:xas}b. The interaction of the $L$-vector with the external magnetic field is mediated by the small but finite ${\bf m}$ which is orientated along the $c$-axis~\cite{Kluczyk23}.

Our main result is the XMCD spectrum for $B=0$~T in Fig.~\ref{fig:xas}a. We find very good agreement between the shapes of the calculated and measured XMCD spectra, with the measured magnitude being about ten times smaller than calculated. We attribute it to presence of magnetic domains with $\bL$'s along the six easy axis directions, see Fig.~\ref{fig:xmcd}c. The calculations assume a single domain sample with an XMCD spectrum $\Delta F(\omega)$. The experimental spectrum, on the other hand, may average over a number of magnetic domains. Importantly, the XMCD spectra 
for $\bL$'s related by $\mathcal{C}_3$ are identical, while those for $-\bL$ have opposite sign following the Onsager
relation $\Delta F(\omega;-\bL, -\bbm)=-\Delta F(\omega;\bL, \bbm)$. 
The shape of the domain-averaged~\footnote{Such averaging has been studied in Ref.~\cite{Kriegner2016} in the case of $T$-even transport phenomena (in MnTe) where the distinction between odd-$p$ and even-$p$ domains was unimportant. The sum then contained only three terms.} spectra 
${\Delta F_{\text{avg}}(\omega)=\left(\sum_{p=0}^5 (-1)^p w_p\right)
    \Delta F(\omega)}$ is identical to the single domain one, 
while the amplitude reflects the imbalance of the weights $w_p$
of the odd-$p$ and even-$p$ domains caused by cooling in the 
magnetic field. The large reduction factor suggests that this imbalance is only moderate. 


Thanks to the localized nature of the core states,  
the X-ray optical conductivity is the sum of contributions from distinct atomic sites:
$\sigma_{\alpha\beta}=\sigma^A_{\alpha\beta}+\sigma^B_{\alpha\beta}$.
For the N\'eel vector $\bL$ in the $c$-plane of \amt, the sublattices $A$ and $B$ are  
connected by the relativistic ${\cal C}_2t_{\sfrac{1}{2}}$ symmetry 
leading to $\sigma^A_{yx}=\sigma^B_{yx}$, $\sigma^A_{zy}=-\sigma^B_{zy}$ and $\sigma^A_{xz}=-\sigma^B_{xz}$. 
Site-resolved contributions to XMCD calculated for $\bbm_A=-\bbm_B$
along $\ca$ are shown in Fig.~\ref{fig:xmcd}ab. 
There are two types of site contributions to XMCD. For light propagating parallel to the local moment, $\bk\parallel\bL$, the individual site contributions are large, and only weakly dependent on the moment orientation, see SM~\cite{sm}. This corresponds to the
common application of XMCD in ferromagnets measuring the projection of $\bbm$ to the light propagation vector $\bk$.
For $\bbm_A=-\bbm_B$ in the $c$-plane it contributes to $\sigma^A_{zy}=-\sigma^B_{zy}$ and $\sigma^A_{xz}=-\sigma^B_{xz}$ elements and thus yields zero net effect. 
The second contribution is found for light propagating perpendicular to the local moments ($\bk\perp \bL$).
This effect is more that an order of magnitude weaker than the $\bk\parallel\bL$ dichroism, but since it contributes to
$\sigma_{yx}$, the identical contributions of the two sublattices add up to a net XMCD. 
Unlike the $\bk\parallel\bL$ dichroism, the $\bk\perp \bL$ contribution strongly depends on the orientation of $\bL$.
It changes sign upon $\bL\rightarrow -\bL$ and vanishes for $\bL\parallel\cb$,
since the corresponding magnetic point group $mmm$ implies that all Hall vector components are zero.

Canting $\bbm_A$ and $\bbm_B$ towards the $c$-axis leads to a finite net magnetization $\bbm$ parallel
to the light propagation vector $\bk$. This gives rise to a per-atom contribution to XMCD that is equal to the $\bk\parallel\bL$ spectra of Fig.~\ref{fig:xmcd} scaled by a factor $|\bbm|/m_{\text{sat}}$, where $m_\text{sat}\approx 5.0~\mu_B$ is the amplitude of the local moment on Mn. The estimated $|\bbm|$ of $10^{-3}$--$10^{-4}\mu_B$
in zero applied field leads to a factor of $2\times 10^{-4}$ -- $2\times 10^{-5}$, which makes the weak-ferromagnetic contribution negligible compared to the $\bk\perp\bL$ contribution. Since $\mathcal{T}$-related domains have opposite
magnetization $\bbm(-\bL)=-\bbm(\bL)$, the domain averaging does not change this conclusion.

The situation is different in the applied magnetic field of $B=6$~T along the $c$-axis.
The induced moment along the $c$-axis $m_{6\text{T}}\approx 0.1~\mu_B$ can be estimated from bulk susceptibility~\cite{sm,Komatsubara1963} and leads to a scaling factor $m_{6\text{T}}/m_\text{sat}\approx 1/50$.
Unlike the canting in zero field, the field-induced canting is the same for all domains, see Fig.~\ref{fig:xmcd}d and thus insensitive
to the domain structure. This allows us to compare not only the spectral shapes, but also the magnitudes
of the calculated and measured in-field XMCD spectra, see Fig.~\ref{fig:xas}. 
The good agreement demonstrates the predictive capability of the present theory not only for the shape, 
but also for the magnitude of the XMCD signal.

\begin{figure}[t]
\includegraphics[width=\columnwidth]{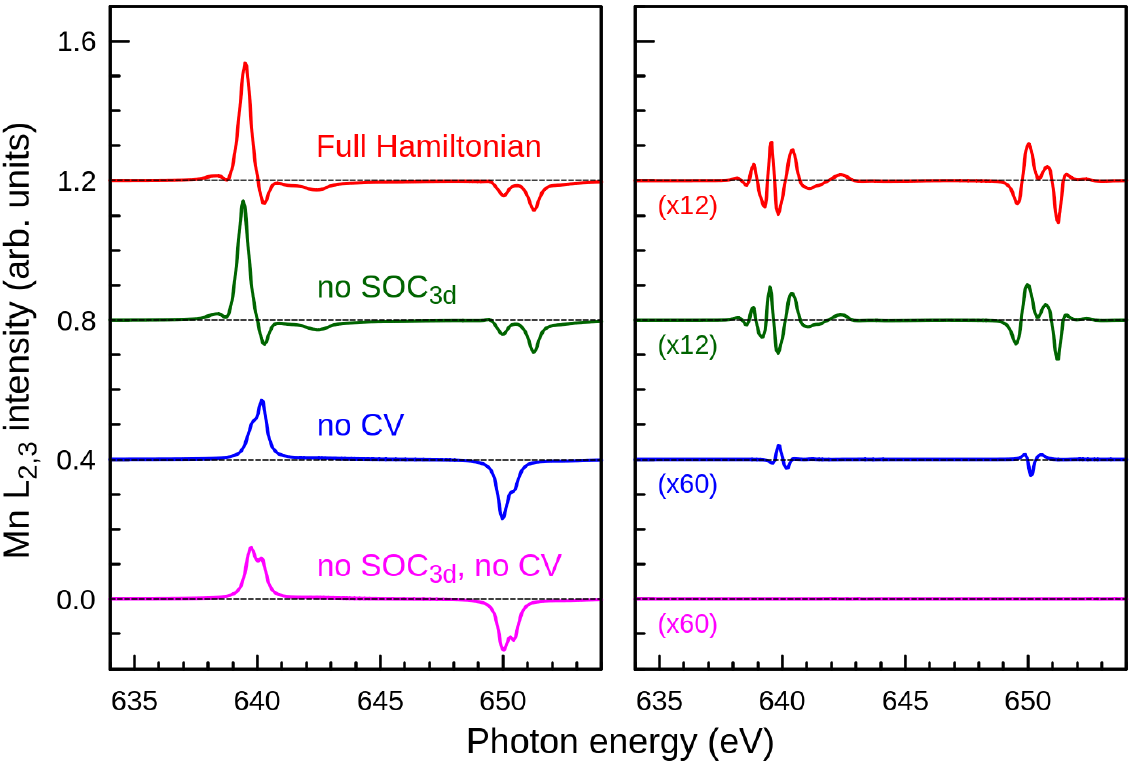}
    \caption{The single-site XMCD for $\bk \parallel {\bf L}$ (left) and $\bk \perp {\bf L}$ (right) for models (i)-(iv) described in the text. The XMCD intensities are magnified by a factor indicated in panels. 
    }
\label{fig:analysis}
\end{figure}

While XMCD and AHE follow the same symmetry rules, they may originate from different terms in the Hamiltonian and thus have different magnitudes or scale differently upon changing the material parameters. An example of such a behavior is different magnitudes of MCD in optical and X-ray regimes 
related to their origin in valence-band and core-level
SOC, respectively. 

For $\bL\parallel\ca$ in \amt, and the corresponding net XMCD given by $\sigma_{yx}$, it is instructive to analyze the single-site XMCD contributions given by $\sigma^{A(B)}_{yx}$ ($\bk\perp \bL$) and $\sigma^{A(B)}_{zy}$ ($\bk\parallel \bL$), although the sum over the two sublattices of the latter vanishes as discussed above. In Fig.~\ref{fig:analysis} we consider four different settings:
i) full Hamiltonian, ii) no valence $3d$ SOC, iii) no core-valence (CV) interaction beyond monopole (e.g. no spin exchange), iv) a combination of (ii) and (iii). 

For $\bk \parallel \bL$, the valence $3d$ SOC plays a negligible role. The core-valence multipole interaction
changes the shape of the single-site XMCD spectra, but does not change the magnitude of the effect.
For $\bk \perp \bL$, we find a moderate difference between (i) and (ii). The single-site XMCD signal is
substantially suppressed in (iii) and vanished completely in (iv). This shows that the role of the valence
$3d$ SOC  is marginal, while the core-valence multipole interaction is crucial for XMCD in \amt. 

The order of magnitude difference between the single-site XMCD for  $\bk \parallel \bL$ and $\bk \perp \bL$, as well as their qualitatively different behavior in case (iv), suggest a different microscopic origin. The single-site XMCD for $\bk \parallel \bL$ corresponds to the conventional XMCD known from ferromagnets, which arises from the spin polarization
of the valence states and SOC in the core states while the role of valence $3d$ SOC and core-valence
interaction is marginal~\footnote{The core-valence interaction changes the shape of XMCD, primarily to due change of XAS.}.

To microscopically understand the XMCD for $\bk \perp \bL$ in \amt, we have to answer
the question: {\it What is the symmetry origin of vanishing of XMCD in case (iv)?}
We use the atomic model for sake of simplicity, nevertheless, the same symmetry arguments apply to the full Hamiltonian of \amt. The terms corresponding to case (iv) are included in 
\begin{equation}
\label{eq:h0}
    \begin{split}
     \hH^0_{\text{at}}&=\epsilon_c\sum_{m,\sigma} \hp^\dagger_{m\sigma}\hp^{\phantom\dagger}_{m\sigma}
        +\!\!\sum_{m,m,\sigma,\sigma'}\! h^{\text{(2p)}}_{m\sigma,m'\!\sigma'} 
        \hp^\dagger_{m\sigma} \hp^{\phantom\dagger}_{m'\!\sigma'}\\
        &+
        \sum_{m,m',\sigma} h^{(\text{CF})}_{mm'} \hd^\dagger_{m\sigma}\hd^{\phantom\dagger}_{m'\!\sigma}
      + b\sum_{m,\sigma} 
        \hd^\dagger_{m\sigma}\hd^{\phantom\dagger}_{m\,-\!\sigma}\\
        &+ U_{pd} \hn_p \hn_d + \sum_{\substack{i,j,k,l\\ \sigma,\sigma'}}u^{\sigma\sigma'}_{ijkl}
        \hd^\dagger_{i\sigma}\hd^{\dagger}_{j\sigma'}
        \hd^{\phantom\dagger}_{k\sigma'}\hd^{\phantom\dagger}_{l\sigma}.
    \end{split}
\end{equation}
The remaining terms to complete the full atomic-model are
\begin{equation*}
\label{eq:h1}
\begin{split}
    \hH^1_{\text{at}}=&\!\sum_{\substack{i,j,k,l\\ \sigma,\sigma'}}w^{\sigma\sigma'}_{ijkl}
        \hd^\dagger_{i\sigma}\hp^\dagger_{j\sigma'}
        \hd^{\phantom\dagger}_{k\sigma'}
        \hp^{\phantom\dagger}_{l\sigma}
        +\!\!\sum_{\substack{m,m'\\ \sigma,\sigma'}} h^{\text{(3d)}}_{m\sigma,m'\!\sigma'} 
        \hd^\dagger_{m\sigma} \hd^{\phantom\dagger}_{m'\!\sigma'}.
        \\       
        \end{split}
\end{equation*}
Here, the operators $\hp^{\dagger}_{m\sigma}$ and $\hd^{\dagger}_{m\sigma}$, respectively, create an electron in a core ($2p$) and valence ($3d$) orbital with angular momentum projection $m$ and spin projection $\sigma=\pm 1$. 
The magnetic order is represented by a Weiss field $b$~\footnote{In AIM the Weiss field is represented by a spin-dependent bath.}, chosen to point along the $x$-axis, 
without loss of generality. 

The XMCD signal for $\bk \perp \bL$ (which is related to the antisymmetric part of $\sigma_{xy}$)
is obtained by the Fermi's golden rule 
\begin{equation}
\label{eq:F}
\begin{split}
&F_\pm(\omega)=\sum_{f} \left|\expval{f|\hT_{\pm}|i}\right|^2\delta\left(\omega-E_{fi}\right)\\
&
 \hT_{\pm}\equiv\sum_\sigma \hT^\sigma_{\pm}=\sum_{m,\sigma}\Gamma^{\phantom n}_{\pm m} \hd^\dagger_{m\pm 1\sigma}\hp^{\phantom\dagger}_{m\sigma}+H.c.
\end{split}
\end{equation}
where $E_{fi}=E_f-E_i$ is the excitation energy from the state $|i\rangle$
to state $|f\rangle$. The dipole operators $\hT_\pm$,  with
real coefficients $\Gamma_{m}$, describe  
absorption of circularly polarized light propagating along the $c$-axis.
The square of the matrix element in (\ref{eq:F}) has the form
\begin{equation}
\label{eq:Ts}
\left|\expval{\hT_{\pm}}\right|^2\!=\!
\sum_\sigma\left|\expval{\hT^\sigma_{\pm}}\right|^2\!+
\sum_\sigma
\expval{\hT^\sigma_{\pm}}\overline{\expval{\hT^{-\sigma}_{\pm}}}.
\end{equation}
First, we use the local ${\cal C}_3$ rotation symmetry about the $c$-axis to show that the second term in
(\ref{eq:Ts}) does not contribute
to $F_\pm(\omega)$. Consider the
$\mathcal{C}_3$ transformation acting only on the valence orbital indices (valence SOC neglected), and on both the core orbital and spin indices (core SOC included)
~\footnote{${\mathcal{C}_3^{\phantom 1}\hp_{m\sigma}\mathcal{C}^{-1}_3=
e^{-i(2m+\sigma)\tfrac{\pi}{3}}\hp_{m\sigma}}$, 
${\mathcal{C}_3^{\phantom 1}\hd_{m\sigma}\mathcal{C}^{-1}_3=
e^{-im\tfrac{2\pi}{3}}\hd_{m\sigma}}$
}.
This operation commutes with $\hH^0_{\text{at}}$ and thus 
\begin{equation}
\label{eq:gsum}
F_\pm(\omega)=\tfrac{1}{3}\!\!\!\!\!\!\!\!\!\sum_{g \in \{I,\mathcal{C}_3,\mathcal{C}^2_3\}}\sum_{f} \left|\expval{f|g\hT_{\pm}g^{-1}|i}\right|^2\!\delta\left(\omega\!-\!E_{fi}\right).
\end{equation}
The transformation of the dipole operator 
\begin{equation}
\label{eq:C3}
\mathcal{C}^{\phantom 1}_3\hT^\sigma_{\pm}\mathcal{C}^{-1}_3 = \varepsilon \hT_\pm^\sigma e^{-i\sigma\tfrac{\pi}{3}},
\end{equation}
introduces a spin-dependent phase shift 
($\varepsilon$ is an overall phase factor) 
arising from the phase difference between the valence spins (not rotated) and the core spins (rotated).
As a result the crossed spin term in (\ref{eq:Ts}) drops out
upon the summation in (\ref{eq:gsum}). 

Next, we consider the role of $\mathcal{T}$. The $\mathcal{T}$-symmetry is  broken 
by the presence of the Weiss field $b$. However, since the valence spin is coupled neither to the valence orbitals, nor to the core spin or orbitals, the transformation
$\mathcal{T}'\equiv\mathcal{T}\mathcal{C}_2^{S,3d}$
\footnote{${\mathcal{T}'\hd_{m\sigma}\mathcal{T}'^{-1}= (-1)^m \hd_{-\!m-\!\sigma}}$,  
${\mathcal{T}'\hp_{m\sigma}\mathcal{T}'^{-1}= (-1)^{m+\tfrac{\sigma-1}{2}} \hp_{-\!m-\!\sigma}}$}  ($\mathcal{T}$ combined with
a $\mathcal{C}_2^{S,3d}$ rotation of the valence spin) is an anti-unitary symmetry, see SM~\cite{sm} for details, which transforms the 
dipole operators as
\begin{equation}
\label{eq:T'}
\mathcal{T}'\hT^\sigma_{\pm}\mathcal{T}'^{-1} = (-1)^{\tfrac{\sigma-1}{2}} \hT^{-\sigma}_\mp.
\end{equation}
As a result we can replace
\begin{equation}
\label{eq:T'element}
    \expval{f|\hT^\sigma_+|i}\rightarrow(-1)^{\tfrac{\sigma-1}{2}}\overline{\expval{f|\hT^{-\sigma}_-|i}}
\end{equation}
in the sum over eigenstates (\ref{eq:F}), which together with vanishing spin-crossed terms in
(\ref{eq:T'element}), leads to $F_+(\omega)=F_-(\omega)$, and thus
zero XMCD. Note that the same result would be obtained for any $\mathcal{C}_n$ rotation axis
parallel to $\bk$.

Turning on either of the terms in $\hH^1_{\text{at}}$ eliminates the above symmetries, and the 
arguments for zero XMCD break down. However, for $\bL\parallel\cb$, XMCD for the full 
Hamiltonian $\hH^0_{\text{at}}+\hH^1_{\text{at}}$
still vanishes. Because of the presence of a mirror plane $\mathcal{M}$ perpendicular to $\bL$, see Fig.~\ref{fig:geometry}, the Hamiltonian is invariant 
under $m\rightarrow -m$ and $\sigma\rightarrow -\sigma$ transformations for both valence and core orbitals. The $\mathcal{M} \hT_+\mathcal{M}^{-1}=\hT_-$ then implies vanishing of XMCD for $\bL\parallel\cb$. The key role of core-valence interaction beyond monopole is reminiscent of the X-ray magnetic linear dichroism in cubic materials~\cite{Mertins2001,Kunes2003}, in which case it also gives rise to a rich
'wiggly' structure in the spectra.~\footnote{In the non-interacting electron picture
of Refs.~\onlinecite{Mertins2001,Kunes2003} the core-valence interaction beyond monopole
is represented in exchange splitting of the core levels.}


Although both AHE and XMCD are given by the anti-symmetric components of (frequency-dependent) $\sigma_{\alpha\beta}$, they arise from different terms in the Hamiltonian. 
AHE originates from SOC in the valence orbitals, i.e., the same interaction responsible for example for magneto-crystalline anisotropy. The electron-electron interactions tend to play a minor role for AHE~\cite{Li:2020_a}, beyond establishing the magnetic order. XMCD in \amt, on the other hand,
shows little sensitivity to valence SOC, but arises from a combination of core SOC and 
core-valence exchange interaction, which affect the excited state containing a core hole.
Only a minor modification of the XMCD signal is observed when the relativistic effects in the valence orbitals are completely neglected.

In conclusion, we have calculated and measured XAS and XMCD at the Mn $L_{2,3}$ edge
in a prototypical altermagnet \amt. 
The calculated effect follows the same symmetry rules as established for AHE. In \amt, it is present for the circularly polarized light propagating along the $c$-axis and the N\'eel vector in the plane perpendicular to the $c$-axis. Within the plane, the effect vanishes for $\bL\parallel [2\bar{1}\bar{1}0]$ and the other two equivalent in-plane axes. 
The calculations show that, unlike AHE, SOC in the valence orbitals plays only a marginal role for XMCD in \amt. 
The distinct shape of XMCD spectra for light propagating perpendicular to the local moments allows to distinguish the altermagnetic
contribution  to XMCD from the weak-ferromagnetic one arising from the moments, judged negligible in MnTe, as well as from possible
contamination by linear dichroism, see SM~\cite{sm}.
The different microscopic origins imply that AHE and XMCD do not scale with each other, despite following the same symmetry rules. As an optical (X-ray) probe, XMCD has also other complementary merits compared to transport experiments, e.g., by enabling microscopy imaging of the domain structure.

\begin{acknowledgements}
We thank Jakub \v{Z}elezn\'y, Satya Prakash Bommanaboyena and Anna Kauch for discussions and critical reading of the manuscript. This work was supported by 
JSPS KAKENHI Grant Numbers 21K13884, 21H01003, 23K03324, 23H03816, 23H03817
(A.H.), projects P30960-N27, I 4493-N (G.S.) and QUAST-FOR5249 project I 5868-N of the Austrian Science Fund (FWF) and
the project Quantum materials for applications in sustainable technologies (QM4ST), funded as project No. CZ.02.01.01/00/22\_008/0004572 by Programme Johannes Amos Commenius, call Excellent Research (J.K.),
the Czech Academy of Sciences, project LQ100102201 (D.K.) and project 22-22000M of the Czech Science Foundation (GACR), and Johannes Gutenberg University Grant TopDyn (L.\v{S}) We thank Diamond Light Source for the allocation of beamtime on beamline I06 under proposals MM33456 and MM36317.
\end{acknowledgements}


\bibliography{main}

\begin{thebibliography}{58}%
\makeatletter
\providecommand \@ifxundefined [1]{%
 \@ifx{#1\undefined}
}%
\providecommand \@ifnum [1]{%
 \ifnum #1\expandafter \@firstoftwo
 \else \expandafter \@secondoftwo
 \fi
}%
\providecommand \@ifx [1]{%
 \ifx #1\expandafter \@firstoftwo
 \else \expandafter \@secondoftwo
 \fi
}%
\providecommand \natexlab [1]{#1}%
\providecommand \enquote  [1]{``#1''}%
\providecommand \bibnamefont  [1]{#1}%
\providecommand \bibfnamefont [1]{#1}%
\providecommand \citenamefont [1]{#1}%
\providecommand \href@noop [0]{\@secondoftwo}%
\providecommand \href [0]{\begingroup \@sanitize@url \@href}%
\providecommand \@href[1]{\@@startlink{#1}\@@href}%
\providecommand \@@href[1]{\endgroup#1\@@endlink}%
\providecommand \@sanitize@url [0]{\catcode `\\12\catcode `\$12\catcode
  `\&12\catcode `\#12\catcode `\^12\catcode `\_12\catcode `\%12\relax}%
\providecommand \@@startlink[1]{}%
\providecommand \@@endlink[0]{}%
\providecommand \url  [0]{\begingroup\@sanitize@url \@url }%
\providecommand \@url [1]{\endgroup\@href {#1}{\urlprefix }}%
\providecommand \urlprefix  [0]{URL }%
\providecommand \Eprint [0]{\href }%
\providecommand \doibase [0]{http://dx.doi.org/}%
\providecommand \selectlanguage [0]{\@gobble}%
\providecommand \bibinfo  [0]{\@secondoftwo}%
\providecommand \bibfield  [0]{\@secondoftwo}%
\providecommand \translation [1]{[#1]}%
\providecommand \BibitemOpen [0]{}%
\providecommand \bibitemStop [0]{}%
\providecommand \bibitemNoStop [0]{.\EOS\space}%
\providecommand \EOS [0]{\spacefactor3000\relax}%
\providecommand \BibitemShut  [1]{\csname bibitem#1\endcsname}%
\let\auto@bib@innerbib\@empty
\bibitem [{\citenamefont {{\v{S}}mejkal}\ \emph {et~al.}(2022)\citenamefont
  {{\v{S}}mejkal}, \citenamefont {MacDonald}, \citenamefont {Sinova},
  \citenamefont {Nakatsuji},\ and\ \citenamefont {Jungwirth}}]{Smejkal22b}%
  \BibitemOpen
  \bibfield  {author} {\bibinfo {author} {\bibfnamefont {L.}~\bibnamefont
  {{\v{S}}mejkal}}, \bibinfo {author} {\bibfnamefont {A.~H.}\ \bibnamefont
  {MacDonald}}, \bibinfo {author} {\bibfnamefont {J.}~\bibnamefont {Sinova}},
  \bibinfo {author} {\bibfnamefont {S.}~\bibnamefont {Nakatsuji}}, \ and\
  \bibinfo {author} {\bibfnamefont {T.}~\bibnamefont {Jungwirth}},\ }\href
  {\doibase 10.1038/s41578-022-00430-3} {\bibfield  {journal} {\bibinfo
  {journal} {Nat. Rev. Mater.}\ }\textbf {\bibinfo {volume} {7}},\ \bibinfo
  {pages} {482} (\bibinfo {year} {2022})}\BibitemShut {NoStop}%
\bibitem [{\citenamefont {\ifmmode~\check{S}\else \v{S}\fi{}mejkal}\ \emph
  {et~al.}(2022{\natexlab{a}})\citenamefont {\ifmmode~\check{S}\else
  \v{S}\fi{}mejkal}, \citenamefont {Sinova},\ and\ \citenamefont
  {Jungwirth}}]{Smejkal22a}%
  \BibitemOpen
  \bibfield  {author} {\bibinfo {author} {\bibfnamefont {L.}~\bibnamefont
  {\ifmmode~\check{S}\else \v{S}\fi{}mejkal}}, \bibinfo {author} {\bibfnamefont
  {J.}~\bibnamefont {Sinova}}, \ and\ \bibinfo {author} {\bibfnamefont
  {T.}~\bibnamefont {Jungwirth}},\ }\href {\doibase 10.1103/PhysRevX.12.040501}
  {\bibfield  {journal} {\bibinfo  {journal} {Phys. Rev. X}\ }\textbf {\bibinfo
  {volume} {12}},\ \bibinfo {pages} {040501} (\bibinfo {year}
  {2022}{\natexlab{a}})}\BibitemShut {NoStop}%
\bibitem [{\citenamefont {\v{S}mejkal}\ \emph {et~al.}(2020)\citenamefont
  {\v{S}mejkal}, \citenamefont {Gonz\'alez-Hern\'andez}, \citenamefont
  {Jungwirth},\ and\ \citenamefont {Sinova}}]{Smejkal20}%
  \BibitemOpen
  \bibfield  {author} {\bibinfo {author} {\bibfnamefont {L.}~\bibnamefont
  {\v{S}mejkal}}, \bibinfo {author} {\bibfnamefont {R.}~\bibnamefont
  {Gonz\'alez-Hern\'andez}}, \bibinfo {author} {\bibfnamefont {T.}~\bibnamefont
  {Jungwirth}}, \ and\ \bibinfo {author} {\bibfnamefont {J.}~\bibnamefont
  {Sinova}},\ }\href {\doibase 10.1126/sciadv.aaz8809} {\bibfield  {journal}
  {\bibinfo  {journal} {Sci. Adv.}\ }\textbf {\bibinfo {volume} {6}},\ \bibinfo
  {pages} {eaaz8809} (\bibinfo {year} {2020})}\BibitemShut {NoStop}%
\bibitem [{\citenamefont {Samanta}\ \emph {et~al.}(2020)\citenamefont
  {Samanta}, \citenamefont {Ležaić}, \citenamefont {Merte}, \citenamefont
  {Freimuth}, \citenamefont {Blügel},\ and\ \citenamefont
  {Mokrousov}}]{Samanta20}%
  \BibitemOpen
  \bibfield  {author} {\bibinfo {author} {\bibfnamefont {K.}~\bibnamefont
  {Samanta}}, \bibinfo {author} {\bibfnamefont {M.}~\bibnamefont {Ležaić}},
  \bibinfo {author} {\bibfnamefont {M.}~\bibnamefont {Merte}}, \bibinfo
  {author} {\bibfnamefont {F.}~\bibnamefont {Freimuth}}, \bibinfo {author}
  {\bibfnamefont {S.}~\bibnamefont {Blügel}}, \ and\ \bibinfo {author}
  {\bibfnamefont {Y.}~\bibnamefont {Mokrousov}},\ }\href {\doibase
  10.1063/5.0005017} {\bibfield  {journal} {\bibinfo  {journal} {J. Appl.
  Phys.}\ }\textbf {\bibinfo {volume} {127}},\ \bibinfo {pages} {213904}
  (\bibinfo {year} {2020})}\BibitemShut {NoStop}%
\bibitem [{\citenamefont {Naka}\ \emph {et~al.}(2020)\citenamefont {Naka},
  \citenamefont {Hayami}, \citenamefont {Kusunose}, \citenamefont {Yanagi},
  \citenamefont {Motome},\ and\ \citenamefont {Seo}}]{Naka20}%
  \BibitemOpen
  \bibfield  {author} {\bibinfo {author} {\bibfnamefont {M.}~\bibnamefont
  {Naka}}, \bibinfo {author} {\bibfnamefont {S.}~\bibnamefont {Hayami}},
  \bibinfo {author} {\bibfnamefont {H.}~\bibnamefont {Kusunose}}, \bibinfo
  {author} {\bibfnamefont {Y.}~\bibnamefont {Yanagi}}, \bibinfo {author}
  {\bibfnamefont {Y.}~\bibnamefont {Motome}}, \ and\ \bibinfo {author}
  {\bibfnamefont {H.}~\bibnamefont {Seo}},\ }\href {\doibase
  10.1103/PhysRevB.102.075112} {\bibfield  {journal} {\bibinfo  {journal}
  {Phys. Rev. B}\ }\textbf {\bibinfo {volume} {102}},\ \bibinfo {pages}
  {075112} (\bibinfo {year} {2020})}\BibitemShut {NoStop}%
\bibitem [{\citenamefont {Reichlová}\ \emph {et~al.}()\citenamefont
  {Reichlová}, \citenamefont {Seeger}, \citenamefont {González-Hernández},
  \citenamefont {Kounta}, \citenamefont {Schlitz}, \citenamefont {Kriegner},
  \citenamefont {Ritzinger}, \citenamefont {Lammel}, \citenamefont {Leiviskä},
  \citenamefont {Petříček}, \citenamefont {Doležal}, \citenamefont
  {Schmoranzerová}, \citenamefont {Bad'ura}, \citenamefont {Thomas},
  \citenamefont {Baltz}, \citenamefont {Michez}, \citenamefont {Sinova},
  \citenamefont {Goennenwein}, \citenamefont {Jungwirth},\ and\ \citenamefont
  {Šmejkal}}]{Reichlova21}%
  \BibitemOpen
  \bibfield  {author} {\bibinfo {author} {\bibfnamefont {H.}~\bibnamefont
  {Reichlová}}, \bibinfo {author} {\bibfnamefont {R.~L.}\ \bibnamefont
  {Seeger}}, \bibinfo {author} {\bibfnamefont {R.}~\bibnamefont
  {González-Hernández}}, \bibinfo {author} {\bibfnamefont {I.}~\bibnamefont
  {Kounta}}, \bibinfo {author} {\bibfnamefont {R.}~\bibnamefont {Schlitz}},
  \bibinfo {author} {\bibfnamefont {D.}~\bibnamefont {Kriegner}}, \bibinfo
  {author} {\bibfnamefont {P.}~\bibnamefont {Ritzinger}}, \bibinfo {author}
  {\bibfnamefont {M.}~\bibnamefont {Lammel}}, \bibinfo {author} {\bibfnamefont
  {M.}~\bibnamefont {Leiviskä}}, \bibinfo {author} {\bibfnamefont
  {V.}~\bibnamefont {Petříček}}, \bibinfo {author} {\bibfnamefont
  {P.}~\bibnamefont {Doležal}}, \bibinfo {author} {\bibfnamefont
  {E.}~\bibnamefont {Schmoranzerová}}, \bibinfo {author} {\bibfnamefont
  {A.}~\bibnamefont {Bad'ura}}, \bibinfo {author} {\bibfnamefont
  {A.}~\bibnamefont {Thomas}}, \bibinfo {author} {\bibfnamefont
  {V.}~\bibnamefont {Baltz}}, \bibinfo {author} {\bibfnamefont
  {L.}~\bibnamefont {Michez}}, \bibinfo {author} {\bibfnamefont
  {J.}~\bibnamefont {Sinova}}, \bibinfo {author} {\bibfnamefont {S.~T.~B.}\
  \bibnamefont {Goennenwein}}, \bibinfo {author} {\bibfnamefont
  {T.}~\bibnamefont {Jungwirth}}, \ and\ \bibinfo {author} {\bibfnamefont
  {L.}~\bibnamefont {Šmejkal}},\ }\href@noop {} {}\Eprint
  {http://arxiv.org/abs/2012.15651} {arXiv:2012.15651} \BibitemShut {NoStop}%
\bibitem [{\citenamefont {Hayami}\ and\ \citenamefont
  {Kusunose}(2021)}]{Hayami21}%
  \BibitemOpen
  \bibfield  {author} {\bibinfo {author} {\bibfnamefont {S.}~\bibnamefont
  {Hayami}}\ and\ \bibinfo {author} {\bibfnamefont {H.}~\bibnamefont
  {Kusunose}},\ }\href {\doibase 10.1103/PhysRevB.103.L180407} {\bibfield
  {journal} {\bibinfo  {journal} {Phys. Rev. B}\ }\textbf {\bibinfo {volume}
  {103}},\ \bibinfo {pages} {L180407} (\bibinfo {year} {2021})}\BibitemShut
  {NoStop}%
\bibitem [{\citenamefont {Mazin}\ \emph {et~al.}(2021)\citenamefont {Mazin},
  \citenamefont {Koepernik}, \citenamefont {Johannes}, \citenamefont
  {González-Hernández},\ and\ \citenamefont {Šmejkal}}]{Mazin21}%
  \BibitemOpen
  \bibfield  {author} {\bibinfo {author} {\bibfnamefont {I.~I.}\ \bibnamefont
  {Mazin}}, \bibinfo {author} {\bibfnamefont {K.}~\bibnamefont {Koepernik}},
  \bibinfo {author} {\bibfnamefont {M.~D.}\ \bibnamefont {Johannes}}, \bibinfo
  {author} {\bibfnamefont {R.}~\bibnamefont {González-Hernández}}, \ and\
  \bibinfo {author} {\bibfnamefont {L.}~\bibnamefont {Šmejkal}},\ }\href
  {\doibase 10.1073/pnas.2108924118} {\bibfield  {journal} {\bibinfo  {journal}
  {Proc. Natl. Acad. Sci. U.S.A.}\ }\textbf {\bibinfo {volume} {118}},\
  \bibinfo {pages} {e2108924118} (\bibinfo {year} {2021})}\BibitemShut
  {NoStop}%
\bibitem [{\citenamefont {Gonzalez~Betancourt}\ \emph
  {et~al.}(2023)\citenamefont {Gonzalez~Betancourt}, \citenamefont
  {Zub\'a\ifmmode~\check{c}\else \v{c}\fi{}}, \citenamefont
  {Gonzalez-Hernandez}, \citenamefont {Geishendorf}, \citenamefont {\ifmmode
  \check{S}\else \v{S}\fi{}ob\'a\ifmmode~\check{n}\else \v{n}\fi{}},
  \citenamefont {Springholz}, \citenamefont {Olejn\'{\i}k}, \citenamefont
  {\ifmmode~\check{S}\else \v{S}\fi{}mejkal}, \citenamefont {Sinova},
  \citenamefont {Jungwirth}, \citenamefont {Goennenwein}, \citenamefont
  {Thomas}, \citenamefont {Reichlov\'a}, \citenamefont {\ifmmode~\check{Z}\else
  \v{Z}\fi{}elezn\'y},\ and\ \citenamefont {Kriegner}}]{Betancourt23}%
  \BibitemOpen
  \bibfield  {author} {\bibinfo {author} {\bibfnamefont {R.~D.}\ \bibnamefont
  {Gonzalez~Betancourt}}, \bibinfo {author} {\bibfnamefont {J.}~\bibnamefont
  {Zub\'a\ifmmode~\check{c}\else \v{c}\fi{}}}, \bibinfo {author} {\bibfnamefont
  {R.}~\bibnamefont {Gonzalez-Hernandez}}, \bibinfo {author} {\bibfnamefont
  {K.}~\bibnamefont {Geishendorf}}, \bibinfo {author} {\bibfnamefont
  {Z.}~\bibnamefont {\ifmmode \check{S}\else
  \v{S}\fi{}ob\'a\ifmmode~\check{n}\else \v{n}\fi{}}}, \bibinfo {author}
  {\bibfnamefont {G.}~\bibnamefont {Springholz}}, \bibinfo {author}
  {\bibfnamefont {K.}~\bibnamefont {Olejn\'{\i}k}}, \bibinfo {author}
  {\bibfnamefont {L.}~\bibnamefont {\ifmmode~\check{S}\else \v{S}\fi{}mejkal}},
  \bibinfo {author} {\bibfnamefont {J.}~\bibnamefont {Sinova}}, \bibinfo
  {author} {\bibfnamefont {T.}~\bibnamefont {Jungwirth}}, \bibinfo {author}
  {\bibfnamefont {S.~T.~B.}\ \bibnamefont {Goennenwein}}, \bibinfo {author}
  {\bibfnamefont {A.}~\bibnamefont {Thomas}}, \bibinfo {author} {\bibfnamefont
  {H.}~\bibnamefont {Reichlov\'a}}, \bibinfo {author} {\bibfnamefont
  {J.}~\bibnamefont {\ifmmode~\check{Z}\else \v{Z}\fi{}elezn\'y}}, \ and\
  \bibinfo {author} {\bibfnamefont {D.}~\bibnamefont {Kriegner}},\ }\href
  {\doibase 10.1103/PhysRevLett.130.036702} {\bibfield  {journal} {\bibinfo
  {journal} {Phys. Rev. Lett.}\ }\textbf {\bibinfo {volume} {130}},\ \bibinfo
  {pages} {036702} (\bibinfo {year} {2023})}\BibitemShut {NoStop}%
\bibitem [{\citenamefont {Naka}\ \emph {et~al.}(2022)\citenamefont {Naka},
  \citenamefont {Motome},\ and\ \citenamefont {Seo}}]{Naka22}%
  \BibitemOpen
  \bibfield  {author} {\bibinfo {author} {\bibfnamefont {M.}~\bibnamefont
  {Naka}}, \bibinfo {author} {\bibfnamefont {Y.}~\bibnamefont {Motome}}, \ and\
  \bibinfo {author} {\bibfnamefont {H.}~\bibnamefont {Seo}},\ }\href {\doibase
  10.1103/PhysRevB.106.195149} {\bibfield  {journal} {\bibinfo  {journal}
  {Phys. Rev. B}\ }\textbf {\bibinfo {volume} {106}},\ \bibinfo {pages}
  {195149} (\bibinfo {year} {2022})}\BibitemShut {NoStop}%
\bibitem [{\citenamefont {Naka}\ \emph {et~al.}(2019)\citenamefont {Naka},
  \citenamefont {Hayami}, \citenamefont {Kusunose}, \citenamefont {Yanagi},
  \citenamefont {Motome},\ and\ \citenamefont {Seo}}]{Naka19}%
  \BibitemOpen
  \bibfield  {author} {\bibinfo {author} {\bibfnamefont {M.}~\bibnamefont
  {Naka}}, \bibinfo {author} {\bibfnamefont {S.}~\bibnamefont {Hayami}},
  \bibinfo {author} {\bibfnamefont {H.}~\bibnamefont {Kusunose}}, \bibinfo
  {author} {\bibfnamefont {Y.}~\bibnamefont {Yanagi}}, \bibinfo {author}
  {\bibfnamefont {Y.}~\bibnamefont {Motome}}, \ and\ \bibinfo {author}
  {\bibfnamefont {H.}~\bibnamefont {Seo}},\ }\href {\doibase
  10.1038/s41467-019-12229-y} {\bibfield  {journal} {\bibinfo  {journal} {Nat.
  Commun.}\ }\textbf {\bibinfo {volume} {10}},\ \bibinfo {pages} {4305}
  (\bibinfo {year} {2019})}\BibitemShut {NoStop}%
\bibitem [{\citenamefont {Gonz\'alez-Hern\'andez}\ \emph
  {et~al.}(2021)\citenamefont {Gonz\'alez-Hern\'andez}, \citenamefont
  {\ifmmode~\check{S}\else \v{S}\fi{}mejkal}, \citenamefont {V\'yborn\'y},
  \citenamefont {Yahagi}, \citenamefont {Sinova}, \citenamefont {Jungwirth},\
  and\ \citenamefont {\ifmmode~\check{Z}\else
  \v{Z}\fi{}elezn\'y}}]{Gonzalez-Hernandez21}%
  \BibitemOpen
  \bibfield  {author} {\bibinfo {author} {\bibfnamefont {R.}~\bibnamefont
  {Gonz\'alez-Hern\'andez}}, \bibinfo {author} {\bibfnamefont {L.}~\bibnamefont
  {\ifmmode~\check{S}\else \v{S}\fi{}mejkal}}, \bibinfo {author} {\bibfnamefont
  {K.}~\bibnamefont {V\'yborn\'y}}, \bibinfo {author} {\bibfnamefont
  {Y.}~\bibnamefont {Yahagi}}, \bibinfo {author} {\bibfnamefont
  {J.}~\bibnamefont {Sinova}}, \bibinfo {author} {\bibfnamefont {T.~c.~v.}\
  \bibnamefont {Jungwirth}}, \ and\ \bibinfo {author} {\bibfnamefont
  {J.}~\bibnamefont {\ifmmode~\check{Z}\else \v{Z}\fi{}elezn\'y}},\ }\href
  {\doibase 10.1103/PhysRevLett.126.127701} {\bibfield  {journal} {\bibinfo
  {journal} {Phys. Rev. Lett.}\ }\textbf {\bibinfo {volume} {126}},\ \bibinfo
  {pages} {127701} (\bibinfo {year} {2021})}\BibitemShut {NoStop}%
\bibitem [{\citenamefont {Naka}\ \emph {et~al.}(2021)\citenamefont {Naka},
  \citenamefont {Motome},\ and\ \citenamefont {Seo}}]{Naka21}%
  \BibitemOpen
  \bibfield  {author} {\bibinfo {author} {\bibfnamefont {M.}~\bibnamefont
  {Naka}}, \bibinfo {author} {\bibfnamefont {Y.}~\bibnamefont {Motome}}, \ and\
  \bibinfo {author} {\bibfnamefont {H.}~\bibnamefont {Seo}},\ }\href {\doibase
  10.1103/PhysRevB.103.125114} {\bibfield  {journal} {\bibinfo  {journal}
  {Phys. Rev. B}\ }\textbf {\bibinfo {volume} {103}},\ \bibinfo {pages}
  {125114} (\bibinfo {year} {2021})}\BibitemShut {NoStop}%
\bibitem [{\citenamefont {Ma}\ \emph {et~al.}(2021)\citenamefont {Ma},
  \citenamefont {Hu}, \citenamefont {Li}, \citenamefont {Liu}, \citenamefont
  {Yao}, \citenamefont {Jia},\ and\ \citenamefont {Liu}}]{Ma21}%
  \BibitemOpen
  \bibfield  {author} {\bibinfo {author} {\bibfnamefont {H.-Y.}\ \bibnamefont
  {Ma}}, \bibinfo {author} {\bibfnamefont {M.}~\bibnamefont {Hu}}, \bibinfo
  {author} {\bibfnamefont {N.}~\bibnamefont {Li}}, \bibinfo {author}
  {\bibfnamefont {J.}~\bibnamefont {Liu}}, \bibinfo {author} {\bibfnamefont
  {W.}~\bibnamefont {Yao}}, \bibinfo {author} {\bibfnamefont {J.-F.}\
  \bibnamefont {Jia}}, \ and\ \bibinfo {author} {\bibfnamefont
  {J.}~\bibnamefont {Liu}},\ }\href {\doibase 10.1038/s41467-021-23127-7}
  {\bibfield  {journal} {\bibinfo  {journal} {Nat. Commun.}\ }\textbf {\bibinfo
  {volume} {12}},\ \bibinfo {pages} {2846} (\bibinfo {year}
  {2021})}\BibitemShut {NoStop}%
\bibitem [{\citenamefont {\ifmmode~\check{S}\else \v{S}\fi{}mejkal}\ \emph
  {et~al.}(2022{\natexlab{b}})\citenamefont {\ifmmode~\check{S}\else
  \v{S}\fi{}mejkal}, \citenamefont {Hellenes}, \citenamefont
  {Gonz\'alez-Hern\'andez}, \citenamefont {Sinova},\ and\ \citenamefont
  {Jungwirth}}]{Smejkal22c}%
  \BibitemOpen
  \bibfield  {author} {\bibinfo {author} {\bibfnamefont {L.}~\bibnamefont
  {\ifmmode~\check{S}\else \v{S}\fi{}mejkal}}, \bibinfo {author} {\bibfnamefont
  {A.~B.}\ \bibnamefont {Hellenes}}, \bibinfo {author} {\bibfnamefont
  {R.}~\bibnamefont {Gonz\'alez-Hern\'andez}}, \bibinfo {author} {\bibfnamefont
  {J.}~\bibnamefont {Sinova}}, \ and\ \bibinfo {author} {\bibfnamefont
  {T.}~\bibnamefont {Jungwirth}},\ }\href {\doibase 10.1103/PhysRevX.12.011028}
  {\bibfield  {journal} {\bibinfo  {journal} {Phys. Rev. X}\ }\textbf {\bibinfo
  {volume} {12}},\ \bibinfo {pages} {011028} (\bibinfo {year}
  {2022}{\natexlab{b}})}\BibitemShut {NoStop}%
\bibitem [{\citenamefont {\ifmmode~\check{S}\else \v{S}\fi{}mejkal}\ \emph
  {et~al.}(2022{\natexlab{c}})\citenamefont {\ifmmode~\check{S}\else
  \v{S}\fi{}mejkal}, \citenamefont {Sinova},\ and\ \citenamefont
  {Jungwirth}}]{Smejkal22}%
  \BibitemOpen
  \bibfield  {author} {\bibinfo {author} {\bibfnamefont {L.}~\bibnamefont
  {\ifmmode~\check{S}\else \v{S}\fi{}mejkal}}, \bibinfo {author} {\bibfnamefont
  {J.}~\bibnamefont {Sinova}}, \ and\ \bibinfo {author} {\bibfnamefont
  {T.}~\bibnamefont {Jungwirth}},\ }\href {\doibase 10.1103/PhysRevX.12.031042}
  {\bibfield  {journal} {\bibinfo  {journal} {Phys. Rev. X}\ }\textbf {\bibinfo
  {volume} {12}},\ \bibinfo {pages} {031042} (\bibinfo {year}
  {2022}{\natexlab{c}})}\BibitemShut {NoStop}%
\bibitem [{\citenamefont {Ahn}\ \emph {et~al.}(2019)\citenamefont {Ahn},
  \citenamefont {Hariki}, \citenamefont {Lee},\ and\ \citenamefont
  {Kune\ifmmode~\check{s}\else \v{s}\fi{}}}]{Ahn19}%
  \BibitemOpen
  \bibfield  {author} {\bibinfo {author} {\bibfnamefont {K.-H.}\ \bibnamefont
  {Ahn}}, \bibinfo {author} {\bibfnamefont {A.}~\bibnamefont {Hariki}},
  \bibinfo {author} {\bibfnamefont {K.-W.}\ \bibnamefont {Lee}}, \ and\
  \bibinfo {author} {\bibfnamefont {J.}~\bibnamefont
  {Kune\ifmmode~\check{s}\else \v{s}\fi{}}},\ }\href {\doibase
  10.1103/PhysRevB.99.184432} {\bibfield  {journal} {\bibinfo  {journal} {Phys.
  Rev. B}\ }\textbf {\bibinfo {volume} {99}},\ \bibinfo {pages} {184432}
  (\bibinfo {year} {2019})}\BibitemShut {NoStop}%
\bibitem [{\citenamefont {Hayami}\ \emph {et~al.}(2019)\citenamefont {Hayami},
  \citenamefont {Yanagi},\ and\ \citenamefont {Kusunose}}]{Hayami19}%
  \BibitemOpen
  \bibfield  {author} {\bibinfo {author} {\bibfnamefont {S.}~\bibnamefont
  {Hayami}}, \bibinfo {author} {\bibfnamefont {Y.}~\bibnamefont {Yanagi}}, \
  and\ \bibinfo {author} {\bibfnamefont {H.}~\bibnamefont {Kusunose}},\ }\href
  {\doibase 10.7566/JPSJ.88.123702} {\bibfield  {journal} {\bibinfo  {journal}
  {J. Phys. Soc. Jpn.}\ }\textbf {\bibinfo {volume} {88}},\ \bibinfo {pages}
  {123702} (\bibinfo {year} {2019})}\BibitemShut {NoStop}%
\bibitem [{\citenamefont {Yuan}\ \emph {et~al.}(2020)\citenamefont {Yuan},
  \citenamefont {Wang}, \citenamefont {Luo}, \citenamefont {Rashba},\ and\
  \citenamefont {Zunger}}]{Yuan20}%
  \BibitemOpen
  \bibfield  {author} {\bibinfo {author} {\bibfnamefont {L.-D.}\ \bibnamefont
  {Yuan}}, \bibinfo {author} {\bibfnamefont {Z.}~\bibnamefont {Wang}}, \bibinfo
  {author} {\bibfnamefont {J.-W.}\ \bibnamefont {Luo}}, \bibinfo {author}
  {\bibfnamefont {E.~I.}\ \bibnamefont {Rashba}}, \ and\ \bibinfo {author}
  {\bibfnamefont {A.}~\bibnamefont {Zunger}},\ }\href {\doibase
  10.1103/PhysRevB.102.014422} {\bibfield  {journal} {\bibinfo  {journal}
  {Phys. Rev. B}\ }\textbf {\bibinfo {volume} {102}},\ \bibinfo {pages}
  {014422} (\bibinfo {year} {2020})}\BibitemShut {NoStop}%
\bibitem [{\citenamefont {Yuan}\ \emph {et~al.}(2021)\citenamefont {Yuan},
  \citenamefont {Wang}, \citenamefont {Luo},\ and\ \citenamefont
  {Zunger}}]{Yuan21}%
  \BibitemOpen
  \bibfield  {author} {\bibinfo {author} {\bibfnamefont {L.-D.}\ \bibnamefont
  {Yuan}}, \bibinfo {author} {\bibfnamefont {Z.}~\bibnamefont {Wang}}, \bibinfo
  {author} {\bibfnamefont {J.-W.}\ \bibnamefont {Luo}}, \ and\ \bibinfo
  {author} {\bibfnamefont {A.}~\bibnamefont {Zunger}},\ }\href {\doibase
  10.1103/PhysRevMaterials.5.014409} {\bibfield  {journal} {\bibinfo  {journal}
  {Phys. Rev. Mater.}\ }\textbf {\bibinfo {volume} {5}},\ \bibinfo {pages}
  {014409} (\bibinfo {year} {2021})}\BibitemShut {NoStop}%
\bibitem [{\citenamefont {Hayami}\ \emph {et~al.}(2020)\citenamefont {Hayami},
  \citenamefont {Yanagi},\ and\ \citenamefont {Kusunose}}]{Hayami20}%
  \BibitemOpen
  \bibfield  {author} {\bibinfo {author} {\bibfnamefont {S.}~\bibnamefont
  {Hayami}}, \bibinfo {author} {\bibfnamefont {Y.}~\bibnamefont {Yanagi}}, \
  and\ \bibinfo {author} {\bibfnamefont {H.}~\bibnamefont {Kusunose}},\ }\href
  {\doibase 10.1103/PhysRevB.102.144441} {\bibfield  {journal} {\bibinfo
  {journal} {Phys. Rev. B}\ }\textbf {\bibinfo {volume} {102}},\ \bibinfo
  {pages} {144441} (\bibinfo {year} {2020})}\BibitemShut {NoStop}%
\bibitem [{\citenamefont {Liu}\ \emph {et~al.}(2022)\citenamefont {Liu},
  \citenamefont {Li}, \citenamefont {Han}, \citenamefont {Wan},\ and\
  \citenamefont {Liu}}]{Liu22}%
  \BibitemOpen
  \bibfield  {author} {\bibinfo {author} {\bibfnamefont {P.}~\bibnamefont
  {Liu}}, \bibinfo {author} {\bibfnamefont {J.}~\bibnamefont {Li}}, \bibinfo
  {author} {\bibfnamefont {J.}~\bibnamefont {Han}}, \bibinfo {author}
  {\bibfnamefont {X.}~\bibnamefont {Wan}}, \ and\ \bibinfo {author}
  {\bibfnamefont {Q.}~\bibnamefont {Liu}},\ }\href {\doibase
  10.1103/PhysRevX.12.021016} {\bibfield  {journal} {\bibinfo  {journal} {Phys.
  Rev. X}\ }\textbf {\bibinfo {volume} {12}},\ \bibinfo {pages} {021016}
  (\bibinfo {year} {2022})}\BibitemShut {NoStop}%
\bibitem [{\citenamefont {Yang}\ \emph {et~al.}()\citenamefont {Yang},
  \citenamefont {Liu},\ and\ \citenamefont {Fang}}]{Jian23}%
  \BibitemOpen
  \bibfield  {author} {\bibinfo {author} {\bibfnamefont {J.}~\bibnamefont
  {Yang}}, \bibinfo {author} {\bibfnamefont {Z.-X.}\ \bibnamefont {Liu}}, \
  and\ \bibinfo {author} {\bibfnamefont {C.}~\bibnamefont {Fang}},\ }\href@noop
  {} {}\Eprint {http://arxiv.org/abs/2105.12738} {arXiv:2105.12738}
  \BibitemShut {NoStop}%
\bibitem [{\citenamefont {Krempask\'y}\ \emph {et~al.}(2023)\citenamefont
  {Krempask\'y}, \citenamefont {\v{S}mejkal}, \citenamefont {D'Souza},
  \citenamefont {Hajlaoui}, \citenamefont {Springholz}, \citenamefont
  {Uhl\'{i}\v{r}ov\'a}, \citenamefont {Alarab}, \citenamefont {Constantinou},
  \citenamefont {Strokov}, \citenamefont {Usanov}, \citenamefont {Pudelko},
  \citenamefont {Gonz\'alez-Hern\'andez}, \citenamefont {Birk~Hellenes},
  \citenamefont {Jansa}, \citenamefont {Reichlov\'a'}, \citenamefont
  {\v{S}ob\'a\v{n}}, \citenamefont {Gonzalez~Betancourt}, \citenamefont
  {Wadley}, \citenamefont {Sinova}, \citenamefont {Kriegner}, \citenamefont
  {Min\'ar}, \citenamefont {Dil},\ and\ \citenamefont
  {Jungwirth}}]{Krempasky23}%
  \BibitemOpen
  \bibfield  {author} {\bibinfo {author} {\bibfnamefont {J.}~\bibnamefont
  {Krempask\'y}}, \bibinfo {author} {\bibfnamefont {L.}~\bibnamefont
  {\v{S}mejkal}}, \bibinfo {author} {\bibfnamefont {S.~W.}\ \bibnamefont
  {D'Souza}}, \bibinfo {author} {\bibfnamefont {M.}~\bibnamefont {Hajlaoui}},
  \bibinfo {author} {\bibfnamefont {G.}~\bibnamefont {Springholz}}, \bibinfo
  {author} {\bibfnamefont {K.}~\bibnamefont {Uhl\'{i}\v{r}ov\'a}}, \bibinfo
  {author} {\bibfnamefont {F.}~\bibnamefont {Alarab}}, \bibinfo {author}
  {\bibfnamefont {P.~C.}\ \bibnamefont {Constantinou}}, \bibinfo {author}
  {\bibfnamefont {V.}~\bibnamefont {Strokov}}, \bibinfo {author} {\bibfnamefont
  {D.}~\bibnamefont {Usanov}}, \bibinfo {author} {\bibfnamefont
  {W.}~\bibnamefont {Pudelko}}, \bibinfo {author} {\bibfnamefont
  {R.}~\bibnamefont {Gonz\'alez-Hern\'andez}}, \bibinfo {author} {\bibfnamefont
  {A.}~\bibnamefont {Birk~Hellenes}}, \bibinfo {author} {\bibfnamefont
  {Z.}~\bibnamefont {Jansa}}, \bibinfo {author} {\bibfnamefont
  {H.}~\bibnamefont {Reichlov\'a'}}, \bibinfo {author} {\bibfnamefont
  {Z.}~\bibnamefont {\v{S}ob\'a\v{n}}}, \bibinfo {author} {\bibfnamefont
  {R.~D.}\ \bibnamefont {Gonzalez~Betancourt}}, \bibinfo {author}
  {\bibfnamefont {P.}~\bibnamefont {Wadley}}, \bibinfo {author} {\bibfnamefont
  {J.}~\bibnamefont {Sinova}}, \bibinfo {author} {\bibfnamefont
  {D.}~\bibnamefont {Kriegner}}, \bibinfo {author} {\bibfnamefont
  {J.}~\bibnamefont {Min\'ar}}, \bibinfo {author} {\bibfnamefont {J.~H.}\
  \bibnamefont {Dil}}, \ and\ \bibinfo {author} {\bibfnamefont
  {T.}~\bibnamefont {Jungwirth}},\ }\href {https://arxiv.org/abs/2308.10681}
  {\bibfield  {journal} {\bibinfo  {journal} {arxiv:2308.10681}\ } (\bibinfo
  {year} {2023})}\BibitemShut {NoStop}%
\bibitem [{\citenamefont {Feng}\ \emph {et~al.}(2022)\citenamefont {Feng},
  \citenamefont {Zhou}, \citenamefont {{\v{S}}mejkal}, \citenamefont {Wu},
  \citenamefont {Zhu}, \citenamefont {Guo}, \citenamefont
  {Gonz{\'a}lez-Hern{\'a}ndez}, \citenamefont {Wang}, \citenamefont {Yan},
  \citenamefont {Qin}, \citenamefont {Zhang}, \citenamefont {Wu}, \citenamefont
  {Chen}, \citenamefont {Meng}, \citenamefont {Liu}, \citenamefont {Xia},
  \citenamefont {Sinova}, \citenamefont {Jungwirth},\ and\ \citenamefont
  {Liu}}]{Feng22}%
  \BibitemOpen
  \bibfield  {author} {\bibinfo {author} {\bibfnamefont {Z.}~\bibnamefont
  {Feng}}, \bibinfo {author} {\bibfnamefont {X.}~\bibnamefont {Zhou}}, \bibinfo
  {author} {\bibfnamefont {L.}~\bibnamefont {{\v{S}}mejkal}}, \bibinfo {author}
  {\bibfnamefont {L.}~\bibnamefont {Wu}}, \bibinfo {author} {\bibfnamefont
  {Z.}~\bibnamefont {Zhu}}, \bibinfo {author} {\bibfnamefont {H.}~\bibnamefont
  {Guo}}, \bibinfo {author} {\bibfnamefont {R.}~\bibnamefont
  {Gonz{\'a}lez-Hern{\'a}ndez}}, \bibinfo {author} {\bibfnamefont
  {X.}~\bibnamefont {Wang}}, \bibinfo {author} {\bibfnamefont {H.}~\bibnamefont
  {Yan}}, \bibinfo {author} {\bibfnamefont {P.}~\bibnamefont {Qin}}, \bibinfo
  {author} {\bibfnamefont {X.}~\bibnamefont {Zhang}}, \bibinfo {author}
  {\bibfnamefont {H.}~\bibnamefont {Wu}}, \bibinfo {author} {\bibfnamefont
  {H.}~\bibnamefont {Chen}}, \bibinfo {author} {\bibfnamefont {Z.}~\bibnamefont
  {Meng}}, \bibinfo {author} {\bibfnamefont {L.}~\bibnamefont {Liu}}, \bibinfo
  {author} {\bibfnamefont {Z.}~\bibnamefont {Xia}}, \bibinfo {author}
  {\bibfnamefont {J.}~\bibnamefont {Sinova}}, \bibinfo {author} {\bibfnamefont
  {T.}~\bibnamefont {Jungwirth}}, \ and\ \bibinfo {author} {\bibfnamefont
  {Z.}~\bibnamefont {Liu}},\ }\href {\doibase 10.1038/s41928-022-00866-z}
  {\bibfield  {journal} {\bibinfo  {journal} {Nat. Electron.}\ }\textbf
  {\bibinfo {volume} {5}},\ \bibinfo {pages} {735} (\bibinfo {year}
  {2022})}\BibitemShut {NoStop}%
\bibitem [{\citenamefont {Shao}\ \emph {et~al.}(2021)\citenamefont {Shao},
  \citenamefont {Zhang}, \citenamefont {Li}, \citenamefont {Eom},\ and\
  \citenamefont {Tsymbal}}]{Shao21}%
  \BibitemOpen
  \bibfield  {author} {\bibinfo {author} {\bibfnamefont {D.-F.}\ \bibnamefont
  {Shao}}, \bibinfo {author} {\bibfnamefont {S.-H.}\ \bibnamefont {Zhang}},
  \bibinfo {author} {\bibfnamefont {M.}~\bibnamefont {Li}}, \bibinfo {author}
  {\bibfnamefont {C.-B.}\ \bibnamefont {Eom}}, \ and\ \bibinfo {author}
  {\bibfnamefont {E.~Y.}\ \bibnamefont {Tsymbal}},\ }\href {\doibase
  10.1038/s41467-021-26915-3} {\bibfield  {journal} {\bibinfo  {journal} {Nat.
  Commun.}\ }\textbf {\bibinfo {volume} {12}},\ \bibinfo {pages} {7061}
  (\bibinfo {year} {2021})}\BibitemShut {NoStop}%
\bibitem [{\citenamefont {Bose}\ \emph {et~al.}(2022)\citenamefont {Bose},
  \citenamefont {Schreiber}, \citenamefont {Jain}, \citenamefont {Shao},
  \citenamefont {Nair}, \citenamefont {Sun}, \citenamefont {Zhang},
  \citenamefont {Muller}, \citenamefont {Tsymbal}, \citenamefont {Schlom},\
  and\ \citenamefont {Ralph}}]{Bose22}%
  \BibitemOpen
  \bibfield  {author} {\bibinfo {author} {\bibfnamefont {A.}~\bibnamefont
  {Bose}}, \bibinfo {author} {\bibfnamefont {N.~J.}\ \bibnamefont {Schreiber}},
  \bibinfo {author} {\bibfnamefont {R.}~\bibnamefont {Jain}}, \bibinfo {author}
  {\bibfnamefont {D.-f.}\ \bibnamefont {Shao}}, \bibinfo {author}
  {\bibfnamefont {H.~P.}\ \bibnamefont {Nair}}, \bibinfo {author}
  {\bibfnamefont {J.}~\bibnamefont {Sun}}, \bibinfo {author} {\bibfnamefont
  {X.~S.}\ \bibnamefont {Zhang}}, \bibinfo {author} {\bibfnamefont {D.~A.}\
  \bibnamefont {Muller}}, \bibinfo {author} {\bibfnamefont {E.~Y.}\
  \bibnamefont {Tsymbal}}, \bibinfo {author} {\bibfnamefont {D.~G.}\
  \bibnamefont {Schlom}}, \ and\ \bibinfo {author} {\bibfnamefont {D.~C.}\
  \bibnamefont {Ralph}},\ }\href {\doibase 10.1038/s41928-022-00758-2}
  {\bibfield  {journal} {\bibinfo  {journal} {Nat. Electron.}\ }\textbf
  {\bibinfo {volume} {5}},\ \bibinfo {pages} {263} (\bibinfo {year} {2022})},\
  \Eprint {http://arxiv.org/abs/2108.09150} {2108.09150} \BibitemShut {NoStop}%
\bibitem [{\citenamefont {Bai}\ \emph {et~al.}(2022)\citenamefont {Bai},
  \citenamefont {Han}, \citenamefont {Feng}, \citenamefont {Zhou},
  \citenamefont {Su}, \citenamefont {Wang}, \citenamefont {Liao}, \citenamefont
  {Zhu}, \citenamefont {Chen}, \citenamefont {Pan}, \citenamefont {Fan},\ and\
  \citenamefont {Song}}]{Bai22}%
  \BibitemOpen
  \bibfield  {author} {\bibinfo {author} {\bibfnamefont {H.}~\bibnamefont
  {Bai}}, \bibinfo {author} {\bibfnamefont {L.}~\bibnamefont {Han}}, \bibinfo
  {author} {\bibfnamefont {X.~Y.}\ \bibnamefont {Feng}}, \bibinfo {author}
  {\bibfnamefont {Y.~J.}\ \bibnamefont {Zhou}}, \bibinfo {author}
  {\bibfnamefont {R.~X.}\ \bibnamefont {Su}}, \bibinfo {author} {\bibfnamefont
  {Q.}~\bibnamefont {Wang}}, \bibinfo {author} {\bibfnamefont {L.~Y.}\
  \bibnamefont {Liao}}, \bibinfo {author} {\bibfnamefont {W.~X.}\ \bibnamefont
  {Zhu}}, \bibinfo {author} {\bibfnamefont {X.~Z.}\ \bibnamefont {Chen}},
  \bibinfo {author} {\bibfnamefont {F.}~\bibnamefont {Pan}}, \bibinfo {author}
  {\bibfnamefont {X.~L.}\ \bibnamefont {Fan}}, \ and\ \bibinfo {author}
  {\bibfnamefont {C.}~\bibnamefont {Song}},\ }\href {\doibase
  10.1103/PhysRevLett.128.197202} {\bibfield  {journal} {\bibinfo  {journal}
  {Phys. Rev. Lett.}\ }\textbf {\bibinfo {volume} {128}},\ \bibinfo {pages}
  {197202} (\bibinfo {year} {2022})}\BibitemShut {NoStop}%
\bibitem [{\citenamefont {Karube}\ \emph {et~al.}(2022)\citenamefont {Karube},
  \citenamefont {Tanaka}, \citenamefont {Sugawara}, \citenamefont {Kadoguchi},
  \citenamefont {Kohda},\ and\ \citenamefont {Nitta}}]{Karube22}%
  \BibitemOpen
  \bibfield  {author} {\bibinfo {author} {\bibfnamefont {S.}~\bibnamefont
  {Karube}}, \bibinfo {author} {\bibfnamefont {T.}~\bibnamefont {Tanaka}},
  \bibinfo {author} {\bibfnamefont {D.}~\bibnamefont {Sugawara}}, \bibinfo
  {author} {\bibfnamefont {N.}~\bibnamefont {Kadoguchi}}, \bibinfo {author}
  {\bibfnamefont {M.}~\bibnamefont {Kohda}}, \ and\ \bibinfo {author}
  {\bibfnamefont {J.}~\bibnamefont {Nitta}},\ }\href {\doibase
  10.1103/PhysRevLett.129.137201} {\bibfield  {journal} {\bibinfo  {journal}
  {Phys. Rev. Lett.}\ }\textbf {\bibinfo {volume} {129}},\ \bibinfo {pages}
  {137201} (\bibinfo {year} {2022})}\BibitemShut {NoStop}%
\bibitem [{\citenamefont {Wimmer}\ \emph {et~al.}(2019)\citenamefont {Wimmer},
  \citenamefont {Mankovsky}, \citenamefont {Min\'ar}, \citenamefont {Yaresko},\
  and\ \citenamefont {Ebert}}]{Wimmer19}%
  \BibitemOpen
  \bibfield  {author} {\bibinfo {author} {\bibfnamefont {S.}~\bibnamefont
  {Wimmer}}, \bibinfo {author} {\bibfnamefont {S.}~\bibnamefont {Mankovsky}},
  \bibinfo {author} {\bibfnamefont {J.}~\bibnamefont {Min\'ar}}, \bibinfo
  {author} {\bibfnamefont {A.~N.}\ \bibnamefont {Yaresko}}, \ and\ \bibinfo
  {author} {\bibfnamefont {H.}~\bibnamefont {Ebert}},\ }\href {\doibase
  10.1103/PhysRevB.100.214429} {\bibfield  {journal} {\bibinfo  {journal}
  {Phys. Rev. B}\ }\textbf {\bibinfo {volume} {100}},\ \bibinfo {pages}
  {214429} (\bibinfo {year} {2019})}\BibitemShut {NoStop}%
\bibitem [{\citenamefont {Kimata}\ \emph {et~al.}(2021)\citenamefont {Kimata},
  \citenamefont {Sasabe}, \citenamefont {Kurita}, \citenamefont {Yamasaki},
  \citenamefont {Tabata}, \citenamefont {Yokoyama}, \citenamefont {Kotani},
  \citenamefont {Ikhlas}, \citenamefont {Tomita}, \citenamefont {Amemiya},
  \citenamefont {Nojiri}, \citenamefont {Nakatsuji}, \citenamefont {Koretsune},
  \citenamefont {Nakao}, \citenamefont {Arima},\ and\ \citenamefont
  {Nakamura}}]{Kimata2021}%
  \BibitemOpen
  \bibfield  {author} {\bibinfo {author} {\bibfnamefont {M.}~\bibnamefont
  {Kimata}}, \bibinfo {author} {\bibfnamefont {N.}~\bibnamefont {Sasabe}},
  \bibinfo {author} {\bibfnamefont {K.}~\bibnamefont {Kurita}}, \bibinfo
  {author} {\bibfnamefont {Y.}~\bibnamefont {Yamasaki}}, \bibinfo {author}
  {\bibfnamefont {C.}~\bibnamefont {Tabata}}, \bibinfo {author} {\bibfnamefont
  {Y.}~\bibnamefont {Yokoyama}}, \bibinfo {author} {\bibfnamefont
  {Y.}~\bibnamefont {Kotani}}, \bibinfo {author} {\bibfnamefont
  {M.}~\bibnamefont {Ikhlas}}, \bibinfo {author} {\bibfnamefont
  {T.}~\bibnamefont {Tomita}}, \bibinfo {author} {\bibfnamefont
  {K.}~\bibnamefont {Amemiya}}, \bibinfo {author} {\bibfnamefont
  {H.}~\bibnamefont {Nojiri}}, \bibinfo {author} {\bibfnamefont
  {S.}~\bibnamefont {Nakatsuji}}, \bibinfo {author} {\bibfnamefont
  {T.}~\bibnamefont {Koretsune}}, \bibinfo {author} {\bibfnamefont
  {H.}~\bibnamefont {Nakao}}, \bibinfo {author} {\bibfnamefont {T.-h.}\
  \bibnamefont {Arima}}, \ and\ \bibinfo {author} {\bibfnamefont
  {T.}~\bibnamefont {Nakamura}},\ }\href {\doibase 10.1038/s41467-021-25834-7}
  {\bibfield  {journal} {\bibinfo  {journal} {Nat. Commun.}\ }\textbf {\bibinfo
  {volume} {12}},\ \bibinfo {pages} {5582} (\bibinfo {year}
  {2021})}\BibitemShut {NoStop}%
\bibitem [{\citenamefont {van~der Laan}(2021)}]{vdLaan2021}%
  \BibitemOpen
  \bibfield  {author} {\bibinfo {author} {\bibfnamefont {G.}~\bibnamefont
  {van~der Laan}},\ }\href {\doibase 10.1103/PhysRevB.104.094414} {\bibfield
  {journal} {\bibinfo  {journal} {Phys. Rev. B}\ }\textbf {\bibinfo {volume}
  {104}},\ \bibinfo {pages} {094414} (\bibinfo {year} {2021})}\BibitemShut
  {NoStop}%
\bibitem [{\citenamefont {\ifmmode~\check{S}\else \v{S}\fi{}mejkal}\ \emph
  {et~al.}(2017)\citenamefont {\ifmmode~\check{S}\else \v{S}\fi{}mejkal},
  \citenamefont {\ifmmode~\check{Z}\else \v{Z}\fi{}elezn\'y}, \citenamefont
  {Sinova},\ and\ \citenamefont {Jungwirth}}]{Smejkal2017}%
  \BibitemOpen
  \bibfield  {author} {\bibinfo {author} {\bibfnamefont {L.}~\bibnamefont
  {\ifmmode~\check{S}\else \v{S}\fi{}mejkal}}, \bibinfo {author} {\bibfnamefont
  {J.}~\bibnamefont {\ifmmode~\check{Z}\else \v{Z}\fi{}elezn\'y}}, \bibinfo
  {author} {\bibfnamefont {J.}~\bibnamefont {Sinova}}, \ and\ \bibinfo {author}
  {\bibfnamefont {T.}~\bibnamefont {Jungwirth}},\ }\href {\doibase
  10.1103/PhysRevLett.118.106402} {\bibfield  {journal} {\bibinfo  {journal}
  {Phys. Rev. Lett.}\ }\textbf {\bibinfo {volume} {118}},\ \bibinfo {pages}
  {106402} (\bibinfo {year} {2017})}\BibitemShut {NoStop}%
\bibitem [{\citenamefont {Elmers}\ \emph {et~al.}(2020)\citenamefont {Elmers},
  \citenamefont {Chernov}, \citenamefont {D’Souza}, \citenamefont
  {Bommanaboyena}, \citenamefont {Bodnar}, \citenamefont {Medjanik},
  \citenamefont {Babenkov}, \citenamefont {Fedchenko}, \citenamefont
  {Vasilyev}, \citenamefont {Agustsson}, \citenamefont {Schlueter},
  \citenamefont {Gloskovskii}, \citenamefont {Matveyev}, \citenamefont
  {Strocov}, \citenamefont {Skourski}, \citenamefont {\v{S}mejkal},
  \citenamefont {Sinova}, \citenamefont {Min\'ar}, \citenamefont {Kl\"aui},
  \citenamefont {Sch\"onhense},\ and\ \citenamefont {Jourdan}}]{Elmers2020}%
  \BibitemOpen
  \bibfield  {author} {\bibinfo {author} {\bibfnamefont {H.~J.}\ \bibnamefont
  {Elmers}}, \bibinfo {author} {\bibfnamefont {S.~V.}\ \bibnamefont {Chernov}},
  \bibinfo {author} {\bibfnamefont {S.~W.}\ \bibnamefont {D’Souza}}, \bibinfo
  {author} {\bibfnamefont {S.~P.}\ \bibnamefont {Bommanaboyena}}, \bibinfo
  {author} {\bibfnamefont {S.~Y.}\ \bibnamefont {Bodnar}}, \bibinfo {author}
  {\bibfnamefont {K.}~\bibnamefont {Medjanik}}, \bibinfo {author}
  {\bibfnamefont {S.}~\bibnamefont {Babenkov}}, \bibinfo {author}
  {\bibfnamefont {O.}~\bibnamefont {Fedchenko}}, \bibinfo {author}
  {\bibfnamefont {D.}~\bibnamefont {Vasilyev}}, \bibinfo {author}
  {\bibfnamefont {S.~Y.}\ \bibnamefont {Agustsson}}, \bibinfo {author}
  {\bibfnamefont {C.}~\bibnamefont {Schlueter}}, \bibinfo {author}
  {\bibfnamefont {A.}~\bibnamefont {Gloskovskii}}, \bibinfo {author}
  {\bibfnamefont {Y.}~\bibnamefont {Matveyev}}, \bibinfo {author}
  {\bibfnamefont {V.~N.}\ \bibnamefont {Strocov}}, \bibinfo {author}
  {\bibfnamefont {Y.}~\bibnamefont {Skourski}}, \bibinfo {author}
  {\bibfnamefont {L.}~\bibnamefont {\v{S}mejkal}}, \bibinfo {author}
  {\bibfnamefont {J.}~\bibnamefont {Sinova}}, \bibinfo {author} {\bibfnamefont
  {J.}~\bibnamefont {Min\'ar}}, \bibinfo {author} {\bibfnamefont
  {M.}~\bibnamefont {Kl\"aui}}, \bibinfo {author} {\bibfnamefont
  {G.}~\bibnamefont {Sch\"onhense}}, \ and\ \bibinfo {author} {\bibfnamefont
  {M.}~\bibnamefont {Jourdan}},\ }\href {\doibase 10.1021/acsnano.0c08215}
  {\bibfield  {journal} {\bibinfo  {journal} {ACS Nano}\ }\textbf {\bibinfo
  {volume} {14}},\ \bibinfo {pages} {17554} (\bibinfo {year}
  {2020})}\BibitemShut {NoStop}%
\bibitem [{vil()}]{villars}%
  \BibitemOpen
  \href@noop {} {}\bibinfo {note} {{P. Villars and K. Cenzual, MnTe Crystal
  Structure: PAULING FILE Multinaries Edition – 2012, Berlin Heidelberg \&
  Material Phases Data System (MPDS), Switzerland \& National Institute for
  Materials Science (NIMS), Japan}}\BibitemShut {NoStop}%
\bibitem [{\citenamefont {Kriegner}\ \emph {et~al.}(2017)\citenamefont
  {Kriegner}, \citenamefont {Reichlova}, \citenamefont {Grenzer}, \citenamefont
  {Schmidt}, \citenamefont {Ressouche}, \citenamefont {Godinho}, \citenamefont
  {Wagner}, \citenamefont {Martin}, \citenamefont {Shick}, \citenamefont
  {Volobuev}, \citenamefont {Springholz}, \citenamefont {Hol\'y}, \citenamefont
  {Wunderlich}, \citenamefont {Jungwirth},\ and\ \citenamefont
  {V\'yborn\'y}}]{Kriegner2017}%
  \BibitemOpen
  \bibfield  {author} {\bibinfo {author} {\bibfnamefont {D.}~\bibnamefont
  {Kriegner}}, \bibinfo {author} {\bibfnamefont {H.}~\bibnamefont {Reichlova}},
  \bibinfo {author} {\bibfnamefont {J.}~\bibnamefont {Grenzer}}, \bibinfo
  {author} {\bibfnamefont {W.}~\bibnamefont {Schmidt}}, \bibinfo {author}
  {\bibfnamefont {E.}~\bibnamefont {Ressouche}}, \bibinfo {author}
  {\bibfnamefont {J.}~\bibnamefont {Godinho}}, \bibinfo {author} {\bibfnamefont
  {T.}~\bibnamefont {Wagner}}, \bibinfo {author} {\bibfnamefont {S.~Y.}\
  \bibnamefont {Martin}}, \bibinfo {author} {\bibfnamefont {A.~B.}\
  \bibnamefont {Shick}}, \bibinfo {author} {\bibfnamefont {V.~V.}\ \bibnamefont
  {Volobuev}}, \bibinfo {author} {\bibfnamefont {G.}~\bibnamefont
  {Springholz}}, \bibinfo {author} {\bibfnamefont {V.}~\bibnamefont {Hol\'y}},
  \bibinfo {author} {\bibfnamefont {J.}~\bibnamefont {Wunderlich}}, \bibinfo
  {author} {\bibfnamefont {T.}~\bibnamefont {Jungwirth}}, \ and\ \bibinfo
  {author} {\bibfnamefont {K.}~\bibnamefont {V\'yborn\'y}},\ }\href {\doibase
  10.1103/PhysRevB.96.214418} {\bibfield  {journal} {\bibinfo  {journal} {Phys.
  Rev. B}\ }\textbf {\bibinfo {volume} {96}},\ \bibinfo {pages} {214418}
  (\bibinfo {year} {2017})}\BibitemShut {NoStop}%
\bibitem [{\citenamefont {Kluczyk}\ \emph {et~al.}(2023)\citenamefont
  {Kluczyk}, \citenamefont {Gas}, \citenamefont {Grzybowski}, \citenamefont
  {Skupinski}, \citenamefont {Borysiewicz}, \citenamefont {Fas}, \citenamefont
  {Suffczynski}, \citenamefont {Domagala}, \citenamefont {Grasza},
  \citenamefont {Mycielski}, \citenamefont {Baj}, \citenamefont {Ahn},
  \citenamefont {Vyborny}, \citenamefont {Sawicki},\ and\ \citenamefont
  {Gryglas-Borysiewicz}}]{Kluczyk23}%
  \BibitemOpen
  \bibfield  {author} {\bibinfo {author} {\bibfnamefont {K.~P.}\ \bibnamefont
  {Kluczyk}}, \bibinfo {author} {\bibfnamefont {K.}~\bibnamefont {Gas}},
  \bibinfo {author} {\bibfnamefont {M.~J.}\ \bibnamefont {Grzybowski}},
  \bibinfo {author} {\bibfnamefont {P.}~\bibnamefont {Skupinski}}, \bibinfo
  {author} {\bibfnamefont {M.~A.}\ \bibnamefont {Borysiewicz}}, \bibinfo
  {author} {\bibfnamefont {T.}~\bibnamefont {Fas}}, \bibinfo {author}
  {\bibfnamefont {J.}~\bibnamefont {Suffczynski}}, \bibinfo {author}
  {\bibfnamefont {J.~Z.}\ \bibnamefont {Domagala}}, \bibinfo {author}
  {\bibfnamefont {K.}~\bibnamefont {Grasza}}, \bibinfo {author} {\bibfnamefont
  {A.}~\bibnamefont {Mycielski}}, \bibinfo {author} {\bibfnamefont
  {M.}~\bibnamefont {Baj}}, \bibinfo {author} {\bibfnamefont {K.~H.}\
  \bibnamefont {Ahn}}, \bibinfo {author} {\bibfnamefont {K.}~\bibnamefont
  {Vyborny}}, \bibinfo {author} {\bibfnamefont {M.}~\bibnamefont {Sawicki}}, \
  and\ \bibinfo {author} {\bibfnamefont {M.}~\bibnamefont
  {Gryglas-Borysiewicz}},\ }\href {https://arxiv.org/abs/2310.09134} {\bibfield
   {journal} {\bibinfo  {journal} {arxiv:2310.09134}\ } (\bibinfo {year}
  {2023})}\BibitemShut {NoStop}%
\bibitem [{\citenamefont {Hariki}\ \emph {et~al.}(2017)\citenamefont {Hariki},
  \citenamefont {Uozumi},\ and\ \citenamefont {Kune\ifmmode~\check{s}\else
  \v{s}\fi{}}}]{Hariki2017}%
  \BibitemOpen
  \bibfield  {author} {\bibinfo {author} {\bibfnamefont {A.}~\bibnamefont
  {Hariki}}, \bibinfo {author} {\bibfnamefont {T.}~\bibnamefont {Uozumi}}, \
  and\ \bibinfo {author} {\bibfnamefont {J.}~\bibnamefont
  {Kune\ifmmode~\check{s}\else \v{s}\fi{}}},\ }\href {\doibase
  10.1103/PhysRevB.96.045111} {\bibfield  {journal} {\bibinfo  {journal} {Phys.
  Rev. B}\ }\textbf {\bibinfo {volume} {96}},\ \bibinfo {pages} {045111}
  (\bibinfo {year} {2017})}\BibitemShut {NoStop}%
\bibitem [{\citenamefont {Hariki}\ \emph {et~al.}(2018)\citenamefont {Hariki},
  \citenamefont {Winder},\ and\ \citenamefont {Kune\ifmmode~\check{s}\else
  \v{s}\fi{}}}]{Hariki18}%
  \BibitemOpen
  \bibfield  {author} {\bibinfo {author} {\bibfnamefont {A.}~\bibnamefont
  {Hariki}}, \bibinfo {author} {\bibfnamefont {M.}~\bibnamefont {Winder}}, \
  and\ \bibinfo {author} {\bibfnamefont {J.}~\bibnamefont
  {Kune\ifmmode~\check{s}\else \v{s}\fi{}}},\ }\href {\doibase
  10.1103/PhysRevLett.121.126403} {\bibfield  {journal} {\bibinfo  {journal}
  {Phys. Rev. Lett.}\ }\textbf {\bibinfo {volume} {121}},\ \bibinfo {pages}
  {126403} (\bibinfo {year} {2018})}\BibitemShut {NoStop}%
\bibitem [{\citenamefont {Hariki}\ \emph {et~al.}(2020)\citenamefont {Hariki},
  \citenamefont {Winder}, \citenamefont {Uozumi},\ and\ \citenamefont
  {Kune\ifmmode~\check{s}\else \v{s}\fi{}}}]{Hariki20}%
  \BibitemOpen
  \bibfield  {author} {\bibinfo {author} {\bibfnamefont {A.}~\bibnamefont
  {Hariki}}, \bibinfo {author} {\bibfnamefont {M.}~\bibnamefont {Winder}},
  \bibinfo {author} {\bibfnamefont {T.}~\bibnamefont {Uozumi}}, \ and\ \bibinfo
  {author} {\bibfnamefont {J.}~\bibnamefont {Kune\ifmmode~\check{s}\else
  \v{s}\fi{}}},\ }\href {\doibase 10.1103/PhysRevB.101.115130} {\bibfield
  {journal} {\bibinfo  {journal} {Phys. Rev. B}\ }\textbf {\bibinfo {volume}
  {101}},\ \bibinfo {pages} {115130} (\bibinfo {year} {2020})}\BibitemShut
  {NoStop}%
\bibitem [{\citenamefont {Blaha}\ \emph {et~al.}()\citenamefont {Blaha},
  \citenamefont {Schwarz}, \citenamefont {Madsen}, \citenamefont {Kvasnicka},\
  and\ \citenamefont {Luitz}}]{wien2k}%
  \BibitemOpen
  \bibfield  {author} {\bibinfo {author} {\bibfnamefont {P.}~\bibnamefont
  {Blaha}}, \bibinfo {author} {\bibfnamefont {K.}~\bibnamefont {Schwarz}},
  \bibinfo {author} {\bibfnamefont {G.}~\bibnamefont {Madsen}}, \bibinfo
  {author} {\bibfnamefont {D.}~\bibnamefont {Kvasnicka}}, \ and\ \bibinfo
  {author} {\bibfnamefont {J.}~\bibnamefont {Luitz}},\ }\href@noop {} {\emph
  {\bibinfo {title} {WIEN2k, An Augmented Plane Wave + Local Orbitals Program
  for Calculating Crystal Properties (Karlheinz Schwarz, Techn. Universitat
  Wien, Austria, 2001), ISBN 3-9501031-1-2}}}\BibitemShut {NoStop}%
\bibitem [{\citenamefont {Kune\v{s}}\ \emph {et~al.}(2010)\citenamefont
  {Kune\v{s}}, \citenamefont {Arita}, \citenamefont {Wissgott}, \citenamefont
  {Toschi}, \citenamefont {Ikeda},\ and\ \citenamefont {Held}}]{wien2wannier}%
  \BibitemOpen
  \bibfield  {author} {\bibinfo {author} {\bibfnamefont {J.}~\bibnamefont
  {Kune\v{s}}}, \bibinfo {author} {\bibfnamefont {R.}~\bibnamefont {Arita}},
  \bibinfo {author} {\bibfnamefont {P.}~\bibnamefont {Wissgott}}, \bibinfo
  {author} {\bibfnamefont {A.}~\bibnamefont {Toschi}}, \bibinfo {author}
  {\bibfnamefont {H.}~\bibnamefont {Ikeda}}, \ and\ \bibinfo {author}
  {\bibfnamefont {K.}~\bibnamefont {Held}},\ }\href {\doibase
  http://dx.doi.org/10.1016/j.cpc.2010.08.005} {\bibfield  {journal} {\bibinfo
  {journal} {Comput. Phys. Commun.}\ }\textbf {\bibinfo {volume} {181}},\
  \bibinfo {pages} {1888 } (\bibinfo {year} {2010})}\BibitemShut {NoStop}%
\bibitem [{\citenamefont {Mostofi}\ \emph {et~al.}(2014)\citenamefont
  {Mostofi}, \citenamefont {Yates}, \citenamefont {Pizzi}, \citenamefont {Lee},
  \citenamefont {Souza}, \citenamefont {Vanderbilt},\ and\ \citenamefont
  {Marzari}}]{wannier90}%
  \BibitemOpen
  \bibfield  {author} {\bibinfo {author} {\bibfnamefont {A.~A.}\ \bibnamefont
  {Mostofi}}, \bibinfo {author} {\bibfnamefont {J.~R.}\ \bibnamefont {Yates}},
  \bibinfo {author} {\bibfnamefont {G.}~\bibnamefont {Pizzi}}, \bibinfo
  {author} {\bibfnamefont {Y.-S.}\ \bibnamefont {Lee}}, \bibinfo {author}
  {\bibfnamefont {I.}~\bibnamefont {Souza}}, \bibinfo {author} {\bibfnamefont
  {D.}~\bibnamefont {Vanderbilt}}, \ and\ \bibinfo {author} {\bibfnamefont
  {N.}~\bibnamefont {Marzari}},\ }\href {\doibase
  http://dx.doi.org/10.1016/j.cpc.2014.05.003} {\bibfield  {journal} {\bibinfo
  {journal} {Comput. Phys. Commun.}\ }\textbf {\bibinfo {volume} {185}},\
  \bibinfo {pages} {2309 } (\bibinfo {year} {2014})}\BibitemShut {NoStop}%
\bibitem [{\citenamefont {Sato}\ \emph {et~al.}(1999)\citenamefont {Sato},
  \citenamefont {Tanaka}, \citenamefont {Furuta}, \citenamefont {Senba},
  \citenamefont {Okuda}, \citenamefont {Mimura}, \citenamefont {Nakatake},
  \citenamefont {Ueda}, \citenamefont {Taniguchi},\ and\ \citenamefont
  {Jo}}]{Sato99}%
  \BibitemOpen
  \bibfield  {author} {\bibinfo {author} {\bibfnamefont {H.}~\bibnamefont
  {Sato}}, \bibinfo {author} {\bibfnamefont {A.}~\bibnamefont {Tanaka}},
  \bibinfo {author} {\bibfnamefont {A.}~\bibnamefont {Furuta}}, \bibinfo
  {author} {\bibfnamefont {S.}~\bibnamefont {Senba}}, \bibinfo {author}
  {\bibfnamefont {H.}~\bibnamefont {Okuda}}, \bibinfo {author} {\bibfnamefont
  {K.}~\bibnamefont {Mimura}}, \bibinfo {author} {\bibfnamefont
  {M.}~\bibnamefont {Nakatake}}, \bibinfo {author} {\bibfnamefont
  {Y.}~\bibnamefont {Ueda}}, \bibinfo {author} {\bibfnamefont {M.}~\bibnamefont
  {Taniguchi}}, \ and\ \bibinfo {author} {\bibfnamefont {T.}~\bibnamefont
  {Jo}},\ }\href {\doibase 10.1143/JPSJ.68.2132} {\bibfield  {journal}
  {\bibinfo  {journal} {J. Phys. Soc. Jpn.}\ }\textbf {\bibinfo {volume}
  {68}},\ \bibinfo {pages} {2132} (\bibinfo {year} {1999})}\BibitemShut
  {NoStop}%
\bibitem [{\citenamefont {Faria~Junior}\ \emph {et~al.}(2023)\citenamefont
  {Faria~Junior}, \citenamefont {de~Mare}, \citenamefont {Zollner},
  \citenamefont {Ahn}, \citenamefont {Erlingsson}, \citenamefont {van
  Schilfgaarde},\ and\ \citenamefont {V\'yborn\'y}}]{pfjr:23}%
  \BibitemOpen
  \bibfield  {author} {\bibinfo {author} {\bibfnamefont {P.~E.}\ \bibnamefont
  {Faria~Junior}}, \bibinfo {author} {\bibfnamefont {K.~A.}\ \bibnamefont
  {de~Mare}}, \bibinfo {author} {\bibfnamefont {K.}~\bibnamefont {Zollner}},
  \bibinfo {author} {\bibfnamefont {K.-h.}\ \bibnamefont {Ahn}}, \bibinfo
  {author} {\bibfnamefont {S.~I.}\ \bibnamefont {Erlingsson}}, \bibinfo
  {author} {\bibfnamefont {M.}~\bibnamefont {van Schilfgaarde}}, \ and\
  \bibinfo {author} {\bibfnamefont {K.}~\bibnamefont {V\'yborn\'y}},\ }\href
  {\doibase 10.1103/PhysRevB.107.L100417} {\bibfield  {journal} {\bibinfo
  {journal} {Phys. Rev. B}\ }\textbf {\bibinfo {volume} {107}},\ \bibinfo
  {pages} {L100417} (\bibinfo {year} {2023})}\BibitemShut {NoStop}%
\bibitem [{\citenamefont {Yi}\ \emph {et~al.}(2010)\citenamefont {Yi},
  \citenamefont {He},\ and\ \citenamefont {Sun}}]{Jian10}%
  \BibitemOpen
  \bibfield  {author} {\bibinfo {author} {\bibfnamefont {J.}~\bibnamefont
  {Yi}}, \bibinfo {author} {\bibfnamefont {X.}~\bibnamefont {He}}, \ and\
  \bibinfo {author} {\bibfnamefont {Y.}~\bibnamefont {Sun}},\ }\href {\doibase
  https://doi.org/10.1016/j.jallcom.2009.10.218} {\bibfield  {journal}
  {\bibinfo  {journal} {J. Alloys Compd.}\ }\textbf {\bibinfo {volume} {491}},\
  \bibinfo {pages} {436} (\bibinfo {year} {2010})}\BibitemShut {NoStop}%
\bibitem [{sm()}]{sm}%
  \BibitemOpen
  \href@noop {} {}\bibinfo {note} {See Supplementary Material for
  details.}\BibitemShut {Stop}%
\bibitem [{Note1()}]{Note1}%
  \BibitemOpen
  \bibinfo {note} {Such averaging has been studied in Ref.~\cite {Kriegner2016}
  in the case of $T$-even transport phenomena (in MnTe) where the distinction
  between odd-$p$ and even-$p$ domains was unimportant. The sum then contained
  only three terms.}\BibitemShut {Stop}%
\bibitem [{\citenamefont {Komatsubara}\ \emph {et~al.}(1963)\citenamefont
  {Komatsubara}, \citenamefont {Murakami},\ and\ \citenamefont
  {Hirahara}}]{Komatsubara1963}%
  \BibitemOpen
  \bibfield  {author} {\bibinfo {author} {\bibfnamefont {T.}~\bibnamefont
  {Komatsubara}}, \bibinfo {author} {\bibfnamefont {M.}~\bibnamefont
  {Murakami}}, \ and\ \bibinfo {author} {\bibfnamefont {E.}~\bibnamefont
  {Hirahara}},\ }\href {\doibase 10.1143/JPSJ.18.356} {\bibfield  {journal}
  {\bibinfo  {journal} {J. Phys. Soc. Jpn.}\ }\textbf {\bibinfo {volume}
  {18}},\ \bibinfo {pages} {356} (\bibinfo {year} {1963})}\BibitemShut
  {NoStop}%
\bibitem [{Note2()}]{Note2}%
  \BibitemOpen
  \bibinfo {note} {The core-valence interaction changes the shape of XMCD,
  primarily to due change of XAS.}\BibitemShut {Stop}%
\bibitem [{Note3()}]{Note3}%
  \BibitemOpen
  \bibinfo {note} {In AIM the Weiss field is represented by a spin-dependent
  bath.}\BibitemShut {Stop}%
\bibitem [{Note4()}]{Note4}%
  \BibitemOpen
  \bibinfo {note} {${\protect \mathcal {C}_3^{\protect \phantom 1}\protect \hat
  {p}_{m\sigma }\protect \mathcal {C}^{-1}_3= e^{-i(2m+\sigma )\protect \tfrac
  {\pi }{3}}\protect \hat {p}_{m\sigma }}$, ${\protect \mathcal {C}_3^{\protect
  \phantom 1}\protect \hat {d}_{m\sigma }\protect \mathcal {C}^{-1}_3=
  e^{-im\protect \tfrac {2\pi }{3}}\protect \hat {d}_{m\sigma }}$}\BibitemShut
  {NoStop}%
\bibitem [{Note5()}]{Note5}%
  \BibitemOpen
  \bibinfo {note} {${\protect \mathcal {T}'\protect \hat {d}_{m\sigma }\protect
  \mathcal {T}'^{-1}= (-1)^m \protect \hat {d}_{-\protect \!m-\protect \!\sigma
  }}$, ${\protect \mathcal {T}'\protect \hat {p}_{m\sigma }\protect \mathcal
  {T}'^{-1}= (-1)^{m+\protect \tfrac {\sigma -1}{2}} \protect \hat
  {p}_{-\protect \!m-\protect \!\sigma }}$}\BibitemShut {NoStop}%
\bibitem [{\citenamefont {Mertins}\ \emph {et~al.}(2001)\citenamefont
  {Mertins}, \citenamefont {Oppeneer}, \citenamefont
  {Kune\ifmmode~\check{s}\else \v{s}\fi{}}, \citenamefont {Gaupp},
  \citenamefont {Abramsohn},\ and\ \citenamefont {Sch\"afers}}]{Mertins2001}%
  \BibitemOpen
  \bibfield  {author} {\bibinfo {author} {\bibfnamefont {H.-C.}\ \bibnamefont
  {Mertins}}, \bibinfo {author} {\bibfnamefont {P.~M.}\ \bibnamefont
  {Oppeneer}}, \bibinfo {author} {\bibfnamefont {J.}~\bibnamefont
  {Kune\ifmmode~\check{s}\else \v{s}\fi{}}}, \bibinfo {author} {\bibfnamefont
  {A.}~\bibnamefont {Gaupp}}, \bibinfo {author} {\bibfnamefont
  {D.}~\bibnamefont {Abramsohn}}, \ and\ \bibinfo {author} {\bibfnamefont
  {F.}~\bibnamefont {Sch\"afers}},\ }\href {\doibase
  10.1103/PhysRevLett.87.047401} {\bibfield  {journal} {\bibinfo  {journal}
  {Phys. Rev. Lett.}\ }\textbf {\bibinfo {volume} {87}},\ \bibinfo {pages}
  {047401} (\bibinfo {year} {2001})}\BibitemShut {NoStop}%
\bibitem [{\citenamefont {Kune\ifmmode~\check{s}\else \v{s}\fi{}}\ and\
  \citenamefont {Oppeneer}(2003)}]{Kunes2003}%
  \BibitemOpen
  \bibfield  {author} {\bibinfo {author} {\bibfnamefont {J.}~\bibnamefont
  {Kune\ifmmode~\check{s}\else \v{s}\fi{}}}\ and\ \bibinfo {author}
  {\bibfnamefont {P.~M.}\ \bibnamefont {Oppeneer}},\ }\href {\doibase
  10.1103/PhysRevB.67.024431} {\bibfield  {journal} {\bibinfo  {journal} {Phys.
  Rev. B}\ }\textbf {\bibinfo {volume} {67}},\ \bibinfo {pages} {024431}
  (\bibinfo {year} {2003})}\BibitemShut {NoStop}%
\bibitem [{Note6()}]{Note6}%
  \BibitemOpen
  \bibinfo {note} {In the non-interacting electron picture of Refs.~\protect
  \rev@citealp {Mertins2001,Kunes2003} the core-valence interaction beyond
  monopole is represented in exchange splitting of the core
  levels.}\BibitemShut {Stop}%
\bibitem [{\citenamefont {Li}\ and\ \citenamefont
  {Levchenko}(2020)}]{Li:2020_a}%
  \BibitemOpen
  \bibfield  {author} {\bibinfo {author} {\bibfnamefont {S.}~\bibnamefont
  {Li}}\ and\ \bibinfo {author} {\bibfnamefont {A.}~\bibnamefont {Levchenko}},\
  }\href {\doibase 10.1103/PhysRevLett.124.156802} {\bibfield  {journal}
  {\bibinfo  {journal} {Phys. Rev. Lett.}\ }\textbf {\bibinfo {volume} {124}},\
  \bibinfo {pages} {156802} (\bibinfo {year} {2020})}\BibitemShut {NoStop}%
\bibitem [{Kri(2016)}]{Kriegner2016}%
  \BibitemOpen
  \href {\doibase 10.1038/ncomms11623} {\bibfield  {journal} {\bibinfo
  {journal} {Nat. Comm.}\ }\textbf {\bibinfo {volume} {7}},\ \bibinfo {pages}
  {11623} (\bibinfo {year} {2016})}\BibitemShut {NoStop}%
\end{thebibliography}%

\end{document}